\documentclass[aip,jcp,reprint,floatfix]{revtex4-1}

\usepackage{graphicx}
\usepackage{amsmath, amsthm}
\usepackage[dvipsnames]{xcolor}
\usepackage{subfigure}
\usepackage{enumerate}
\usepackage[normalem]{ulem}

\begin{document}

\title{Harnessing entropy to enhance toughness in reversibly crosslinked polymer networks}
\author{Nicholas B. Tito}
	\affiliation{Department of Applied Physics, Eindhoven University of Technology, PO Box 513, 5600 MB, Eindhoven, The Netherlands \\ Institute for Complex Molecular Systems, Eindhoven University of Technology, PO Box 513, 5600 MB, Eindhoven, The Netherlands}
	\email{nicholas.b.tito@gmail.com}
\author{Costantino Creton}
	\affiliation{\'Ecole Sup\'erieure de Physique et de Chimie Industrielles de la Ville de Paris (ESPCI) ParisTech, UMR 7615, 10, Rue Vauquelin, 75231 Paris C\'edex 05, France \\ CNRS, UMR 7615, 10, Rue Vauquelin, 75231 Paris C\'edex 05, France \\ Sorbonne-Universit\'es, Universit\'e Pierre et Marie Curie (UPMC) Universit\'e Paris 06, UMR 7615, 10, Rue Vauquelin, 75231 Paris C\'edex 05, France}
\author{Cornelis Storm}
	\affiliation{Department of Applied Physics, Eindhoven University of Technology, PO Box 513, 5600 MB, Eindhoven, The Netherlands \\ Institute for Complex Molecular Systems, Eindhoven University of Technology, PO Box 513, 5600 MB, Eindhoven, The Netherlands}
\author{Wouter G. Ellenbroek}
	\affiliation{Department of Applied Physics, Eindhoven University of Technology, PO Box 513, 5600 MB, Eindhoven, The Netherlands \\ Institute for Complex Molecular Systems, Eindhoven University of Technology, PO Box 513, 5600 MB, Eindhoven, The Netherlands}

\date{\today}
\begin{abstract}
Reversible crosslinking is a design paradigm for polymeric materials, wherein they are microscopically reinforced with chemical species that form \emph{transient} crosslinks between the polymer chains. Besides the potential for self-healing, recent experimental work suggests that freely diffusing reversible crosslinks in polymer networks, such as gels, can enhance the toughness of the material without substantial change in elasticity. This presents the opportunity for making highly elastic materials that can be strained to a large extent before rupturing. Here, we employ Gaussian chain theory, molecular simulation, and polymer self-consistent field theory for networks to construct an equilibrium picture for how reversible crosslinks can toughen a polymer network without affecting its linear elasticity. Maximisation of polymer entropy drives the reversible crosslinks to bind preferentially near the permanent crosslinks in the network, leading to local molecular reinforcement without significant alteration of the network topology. In equilibrium conditions, permanent crosslinks share effectively the load with neighbouring reversible crosslinks, forming multi-functional crosslink points. The network is thereby globally toughened, while the linear elasticity is left largely unaltered. Practical guidelines are proposed to optimise this design in experiment, along with a discussion of key kinetic and timescale considerations.
\end{abstract}

\maketitle

\section{Introduction}

Materials composed of polymers chemically crosslinked into a network, such as in gels and rubbers, eventually break if strained enough \cite{Lake:1967ec, Ligoure:2013bd, Creton:2017bd, Creton:2016fr, Linga:2015cy}. This is because the polymer chains or crosslinks irreversibly rupture once the local force acting on them becomes too large for their covalent bonds to withstand. Strategies for exploiting reversible bonding and crosslinking for dynamic remodelling, in response to stress or other physical changes, is thus an active focus of study. Recent examples include materials---some biological in nature---that self-heal when damaged \cite{Kim:1983em, Zwaag:2007ue, Wool:2008gc, Stukalin:2013ju, Amaral:2017bi}, dynamically adapt to strain by topology change and reversible bonding \cite{Montarnal:2011cr, Ligoure:2013bd, Kloxin:2013hu, Narita:2013kt, Kean:2014hh, Li:2015bm, Neal:2015ju, Kurniawan:2016hp, Imbernon:2016kb}, or reshape and deform in response to light, temperature, or local chemical environment \cite{White:2015if}. Polymer networks with reversible crosslinks form the basis for many of these exciting new materials, attracting a variety of recent theoretical modelling efforts into systems exhibiting transient inter-polymer, intra-polymer, and telechelic bonding \cite{Hoy:2009gm, Sprakel:2009gs, Indei:2010kl, Nabavi:2015ki, West:2015fe, Gordievskaya:2016es, Amin:2016je, Meng:2016hs, Yu:2016dl, Brighenti:2017ff, Vernerey:2017gt, Oyarzun:2018hz, Oyarzun:2018tk, Parada:2018dk}.

\begin{figure}
	\centering
	\subfigure[]{\includegraphics[width= 0.40\textwidth]{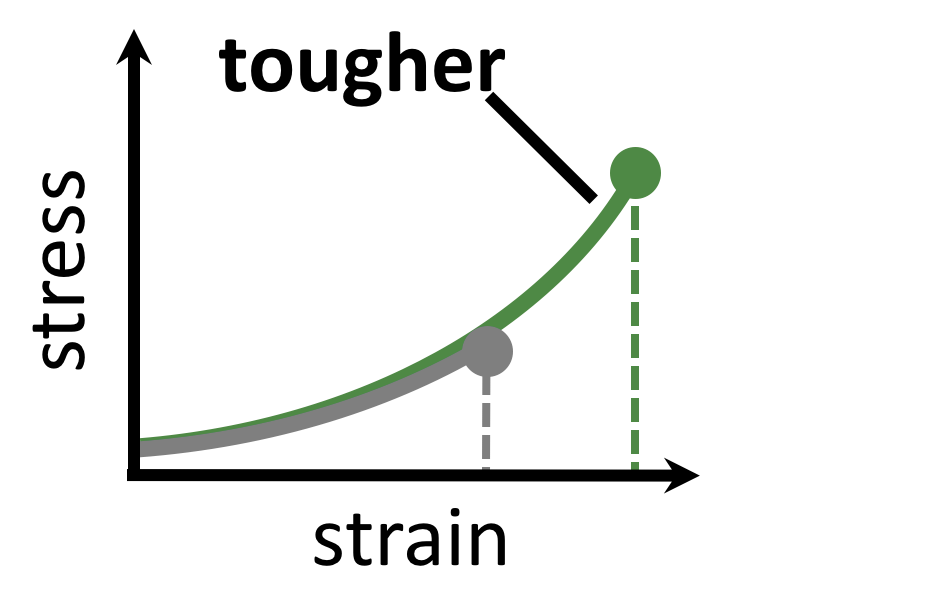}}
	\subfigure[]{\includegraphics[width= 0.40\textwidth]{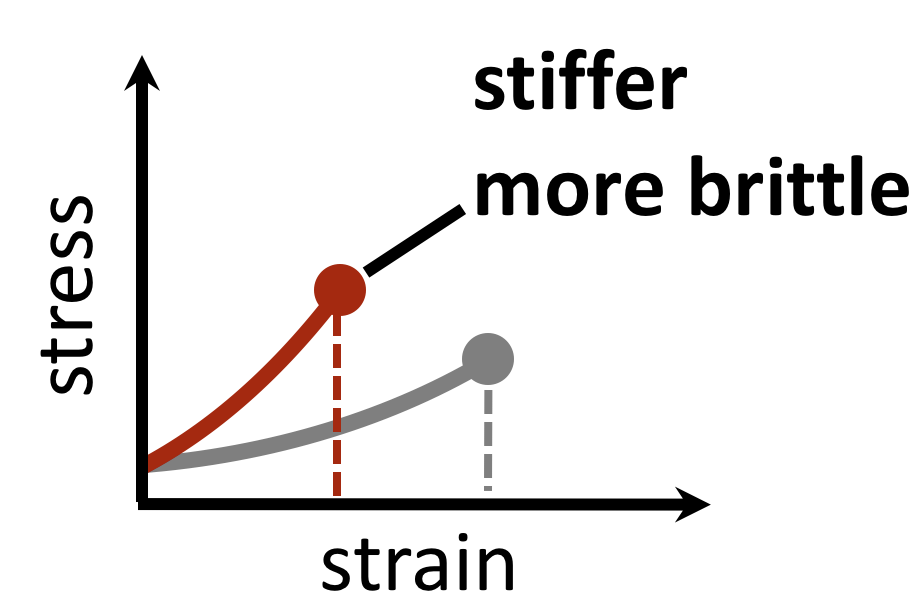}}
	\caption{Stress-strain curves for a hypothetical polymer gels with and without reversible crosslinks. Grey curves in (a) and (b) are gels with no reversible crosslinks. Green curve in (a) is a gel having ``optimally binding'' reversible crosslinks, where the strain rate timescale is longer than the reversible crosslink rearrangment timescale. The coincidence of the green and grey curve up to the failure point is what we term ``having the same elasticity'' throughout this paper. Red curve in (b) is for a gel with strong-binding reversible crosslinks, or for a gel with optimally binding reversible crosslinks when it is rapidly strained. Failure points of the materials are indicated by dots. The red and green polymer gels are imagined to have the same concentration of reversible crosslinking molecules in their solvent phase.}
	\label{fig:MechanicalCartoon}
\end{figure}

Here, attention is focused on how reversible crosslinking can enhance the strength of a network of flexible chains.  Our inspiration comes from recent studies that show that reversible crosslinking can be leveraged to separately tune the toughness of a material, independent of its elasticity.  For example, in two motivating experimental studies by Kean et al.\cite{Kean:2014hh} and Mayumi et al.\cite{Mayumi:2016kr}, reversible crosslinking agents are added into the solvent phase of a permanently crosslinked polymer gel. While the two studies differ in their underlying chemistry, both find clear regimes where the presence of the reversible crosslinks allows the gel to be stretched to a larger extent before it fails---yet remarkably having the same elasticity before failure---compared to the material without the reversible crosslinks. This notion is illustrated by schematic stress-strain curves in Figure \ref{fig:MechanicalCartoon}.  Salient details and observations from these two studies are now summarised, to motivate our objective. 

The experiments by Kean et al. and Mayumi et al. examine gels of permanently crosslinked polymers, comprising poly(4-vinylpyridine)\cite{Kean:2014hh} and polyvinyl alcohol \cite{Mayumi:2016kr} respectively. In Kean et al., the reversible crosslinks are Van Koten-type pincer complexes, composed of two transition metals attached into a small organic scaffold.\cite{Albrecht:2001um} The complex can form one or two coordination bonds, anti-podal to each other, with monomers along the polymer network chains. The complexes are free to diffuse within the solvent phase of the gel, so that they can form and break their coordination bonds at will at \emph{any} position along the polymer chains. The lifetime of the bonds is controlled by altering the chemical structure of the complexes. In Mayumi et al., borate ions act as the reversible crosslinks, which can form up to four transient bonds with the polyvinyl alcohol chains. Note that both reversible crosslinking paradigms have different underlying physics compared to scenarios where the reversible linking units are fixed along the contour lengths of the polymers. \cite{Baxandall:1989kw, Leibler:1991hj, Hackelbusch:2013kq}

Judicious choice of the reversible crosslink binding strength (at fixed loading rate and reversible crosslink concentration) in Kean et al. leads to \emph{toughening} of the gel, without altering its elastic properties.\cite{Kean:2014hh} In this sense, the reversible crosslinks are mechanically ``invisible'', as the authors so put it. When the reversible linkers bind too strongly, then they lead the material to be stiffer and more brittle, with lower toughness.  Conversely, if the reversible crosslinks bind too weakly, then their toughening effect decreases.  In Mayumi et al., the complementary perspective is taken: the polyvinyl alcohol gel is subjected to different loading rates, all having the same borate ion (reversible crosslinker) concentration and binding strength.\cite{Mayumi:2016kr} Here, the largest increase in toughness and smallest change in elasticity, relative to the gel without the reversible crosslinks, occurs in the limit of slow loading rate. Faster loading rates cause the reversible crosslinks to behave more like permanent crosslinks on the timescale of the strain rate, so that the material is apparently stiffer than the native material. 

Reversible crosslinking can allow for somewhat independent control over the toughness of a gel, without an influence on its elasticity. Schematic stress-strain curves for hypothetical polymer gels with and without reversible crosslinks are given in Figure \ref{fig:MechanicalCartoon}. Understanding how reversible crosslinks enact this toughening at the microscale is thus of great value for the development of new gel materials with tailored properties. The subject of this paper is understanding, on theoretical grounds, the optimal design of reversible crosslinks that leads to maximum toughening, with minimal influence on the elasticity of the material.

An oft-discussed suggestion for how free reversible crosslinks invisibly toughen a polymer network is that, through the course of strain, the reversible crosslinks (re-)migrate via diffusion to locations where microscopic damage (rupture) is actively occuring. In doing so, the reversible links are purported to have a significant structural and regenerative influence in ruptured locations, while not having the dynamical longevity to affect the mechanics of undamaged regions. However, a recent theoretical study on a polymer network composed of mobile crosslinks by Mulla et al. (Refs. \citenum{Mulla:2018uh} and \citenum{Mulla:2018ga}) suggests that the mobile links migrate \emph{away} from areas of damage over time, particularly when the length scale of the ruptures grows large. \cite{Foyart:2016hu}

In this paper, a combination of molecular theory and simulation is used to construct a different argument, based on equilibrium thermodynamics, for how free reversible crosslinks can significantly enhance toughness with minimal change of elasticity. 

The starting point for our theory is the classical observation that microscopic topology of a polymer network has a key role in dictating the macroscopic stress-strain behaviour of the material.\cite{Treloar:1942, JAMES:1947hp, Edwards:1969dd, Flory:1985co, Holzl:1997dd} Adding crosslinks or chains to a network leads to a stiffer and less extensible material, due to the larger entropic restoring force and shorter average length of the polymers at the molecular scale. On the other hand, more dilute networks of the same material exhibit a lower modulus. If reversible crosslinks can toughen a polymer gel without changing its elasticity, then the microscopic mechanism of toughening must leave the load-supporting part of the molecular topology of the network mostly unaltered.

To construct an explanation for how this feature arises, we first demonstrate that reversible linkers are entropically biased to form transient connections adjacent to permanent crosslinks in the network, when the reversible linker equilibrium binding constant is weak. Forming a transient bond near a permanent crosslink leads to a smaller configurational entropy penalty for the associated polymer strands, relative to forming a transient bond far from a permanent link. The implication of this selective recruitment is that existing permanent crosslinks are less loaded, while the network topology is not changed.  Casting our arguments in terms of thermodynamic entropy, rather than structural proximity effects, opens the doorway for explicitly quantifying how this clustering competes with the other thermodynamic factors, e.g. reversible linker binding enthalpy and concentration.

Molecular dynamics simulations are used to verify this hypothesis, and to show that ``recruiting'' (clustering) reversible crosslinks around permanent crosslinks leads to local load-sharing and reduced chance of failure of the permanent bond when the polymer network is under strain. Coarse-grained modeling of polymer network failure, using a self-consistent field model, demonstrates that local molecular strengthening of permanent crosslinks in the network leads to an amplified toughening of the polymer network as a whole. After a discussion of kinetic considerations around our equilibrium theory, we put forth practical design guidelines with the aim of illuminating future experimental efforts.

\section{Entropy \& recruitment}

Polymers have an intrinsic entropy, depending upon the number of configurations they can explore at equilibrium.\cite{DeGennes1979, Rubinstein2003} Free polymers have higher entropy than constrained polymers. When permanently crosslinked into a network, polymers have an inherently smaller entropy, yet the polymers still explore a space of configurations at equilibrium which satisfy the fixed connectivity of the network. Their non-trivial entropy can be further reduced by adding additional constraints, such as transient crosslinks.

\begin{figure}
	\centering
	\subfigure[]{\includegraphics[width= 0.48\textwidth]{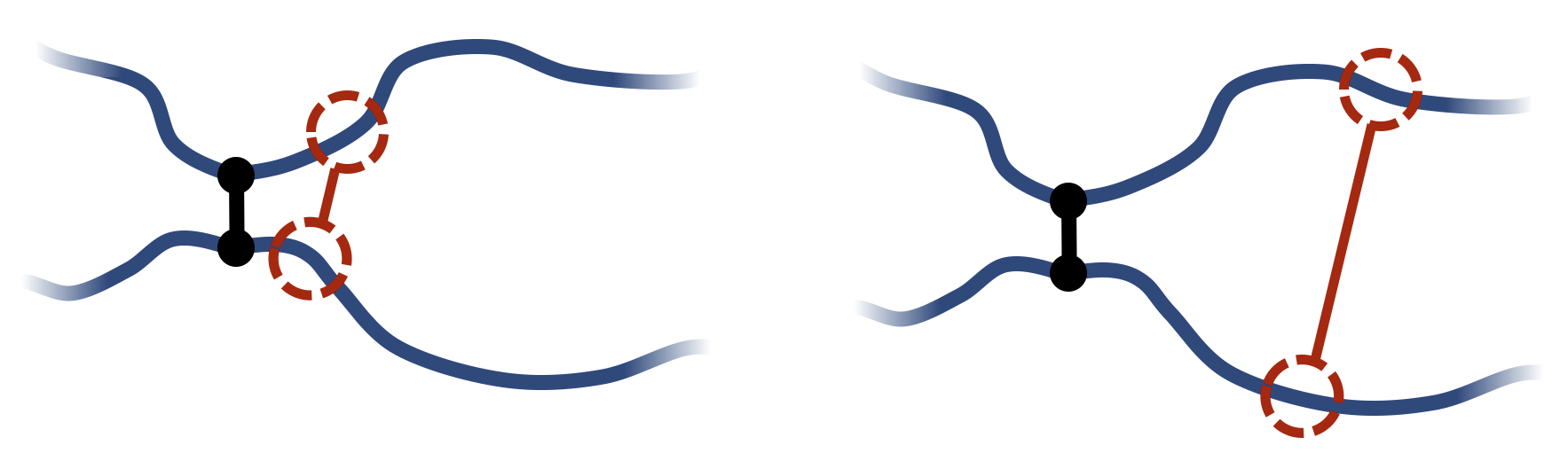}}
	\subfigure[]{\includegraphics[width= 0.48\textwidth]{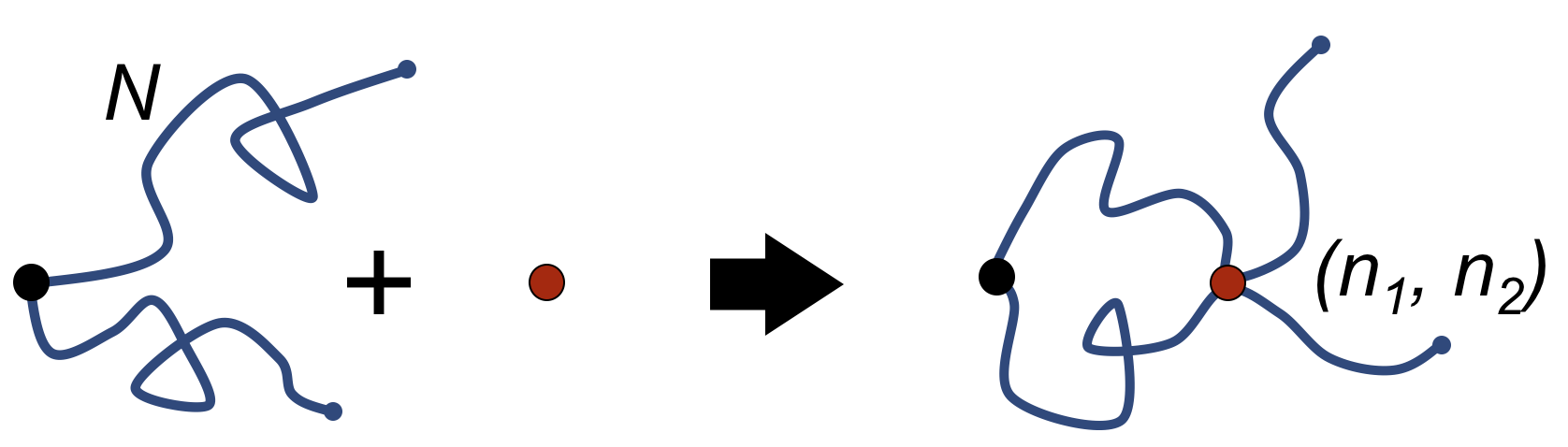}}
	\caption{(a) Illustration of two polymers bound together by a permanent crosslink in a polymer network, and possible choices of binding positions along the polymers for adding a reversible crosslink. Adding the reversible bond close to the permanent crosslink (left) incurs a lower entropy penalty to the polymers compared to if it adds far away (right). (b) Two Gaussian polymers of $N$ segments bound together at the origin by a permanent crosslink (black dot). A reversible crosslink (red dot) binds to segments $(n_1, n_2)$ along the polymers (right).}
	\label{fig:ChainEntropyCartoon}
\end{figure}

Adding a transient bond between two polymers in a network imposes a temporary constraint on the network configuration. This entails an entropic penalty that depends on the position of the bond, relative to the existing crosslinks in the network, as shown in Figure \ref{fig:ChainEntropyCartoon}(a). Within the approximation of ideal Gaussian polymer chains, we can compute the corresponding free energy change exactly. We examine this by considering two Gaussian polymer chains of finite number of segments $N$, each attached to a permanent crosslink by their first segment. The permanent crosslink defines the origin of the system. For the moment, the other two ends of the chains are left untethered.

It is convenient to write the partition function of each polymer using the normalized Gaussian probability density for the end-to-end vector $\mathbf{R}$ of a chain of $N$ segments with Kuhn length $b$
\begin{equation}
    \label{gausschain}
    P(\mathbf{R}, N)=\left(\frac{3}{2\pi N b^2}\right)^{3/2}\exp\left[-\frac{3\mathbf{R}^2}{2Nb^2}\right].
\end{equation}
This way, any linear polymer with an unconstrained end-to-end vector (and even any loopless branched polymer with free ends) will have a partition function
\begin{equation}
	Q=\int P(\mathbf{R}, N) \ \mathrm{d}\mathbf{R} = 1,
\end{equation}
while loops of $n$ segments contribute a factor
\begin{align}
    Q_\mathrm{loop}(n) &= \int P(\mathbf{R}, n) b^3 \delta(\mathbf{R} - \mathbf{R}_o) \ \mathrm{d}\mathbf{R} \nonumber \\
    &= P(0, n) b^3 = \left(\frac{3}{2\pi n}\right)^{3/2},
\end{align}
where $\mathbf{R}_o$ is the fixed origin of the two polymers. The volume element $b^3$ is employed here to yield a dimensionless partition function, however it only contributes a constant additive factor to all free energies henceforth.

\begin{figure}
	\centering
	\includegraphics[width= 0.3\textwidth]{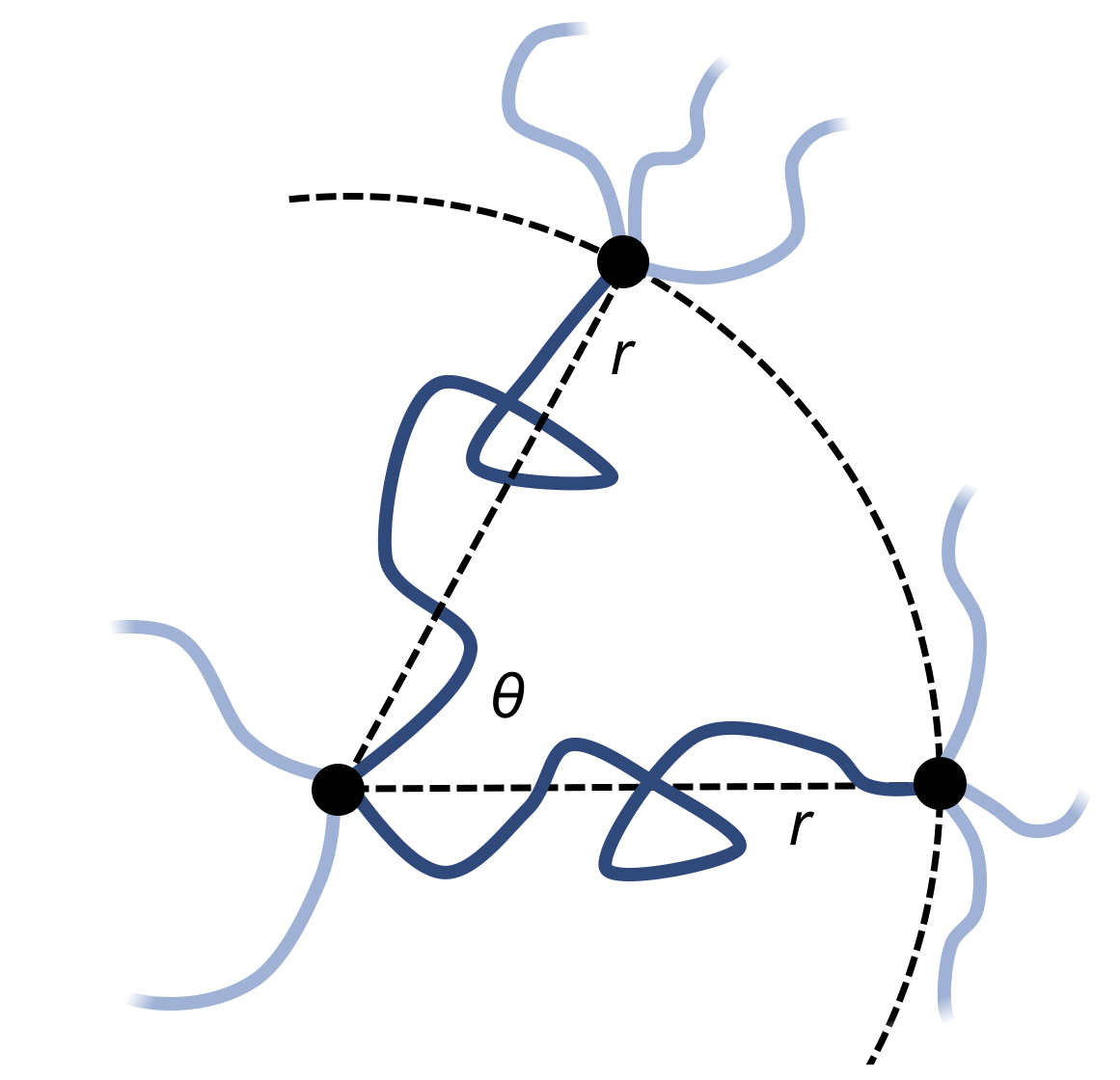}
	\caption{Two polymer chains with two of their endpoints fixed at the origin via a permanent crosslink, and their other endpoints fixed at a distance $r$ from the origin, and at an angle of $\theta$ to each other. Light-coloured chains represents other polymers in the network, attached to the three permanent crosslinks shown.}
	\label{fig:StretchedChainsCartoon}
\end{figure}

Now consider the event sketched in Fig.~\ref{fig:ChainEntropyCartoon}(b). Two Gaussian chains of $N$ segments are connected to each other at the origin. Assuming the other two ends are free, the partition function of this configuration equals unity with our normalization. When a reversible crosslink ties segment $n_1$ from the first chain to segment $n_2$ from the second chain, a loop of length $n_1+n_2\equiv n$ is formed and the partition function is reduced to $Q_\mathrm{loop}(n)$. The change in conformational free energy of the two chains is therefore
\begin{equation}
    \label{eqn:FreeEnergyBind}
    \beta\Delta G(n)=\ln{\left(1/Q_\mathrm{loop}(n)\right)} = \frac{3}{2}\ln\left(\frac{2\pi n}{3}\right),    
\end{equation}
where $\beta = 1/ kT$, with Boltzmann constant $k$ and temperature $T$. 

As it stands, Eq. (\ref{eqn:FreeEnergyBind}) tells us that the polymer network pays a free energy of order $\frac32\ln(n_1+n_2)$ when a reversible crosslink is added that connects site $n_1$ of chain 1 to site $n_2$ of chain 2. In order for this estimate to account for \emph{all} loops of size $n$, the partition function after binding needs to be multiplied by the number of choices one can make for $n_1$ and $n_2$ such that $n=n_1+n_2$, which equals $\min{\left(n - 1, 2N - n - 1\right)}$. This makes the final change in network free energy due to the creation of a loop of size $n$ through a binding event in Fig. \ref{fig:ChainEntropyCartoon}b equal to
\begin{align}
	\beta \Delta G^{\circ}_{\text{poly}} (n; N) = & \frac{3}{2} \ln{\left(\frac{2 \pi n}{3} \right)} \nonumber \\
	& - \ln{\left[\min{\left(n - 1, 2N - n - 1\right)}\right]}
	\label{eqn:FreeEnergyBindFinal}
\end{align}
The degeneracy adds a term that goes like $-\ln (n-1)$, so the overall free energy still grows logarithmically with $n$. This demonstrates that the polymer network loses the least entropy (i.e. gains the least free energy) when the reversible crosslink binds as close to an existing permanent crosslink as possible.  The permanent crosslinks can be said to act as nucleation points for reversible crosslink binding on the polymer network scaffold.

In real networks, the other ends of the chains are not usually free, but tied to the rest of the network at other points. As a related scenario, we now turn to the case where the free ends of the two polymers are constrained to two points, both for simplicity at the same distance $r$ from the permanent crosslink and with the same number of segments $N$. The distance $r$ of the additional attachment points from the permanent crosslink, as well as the angle $\theta$ between them, can be varied as illustrated in Figure \ref{fig:StretchedChainsCartoon}. 

The initial free energy for each chain is identical, since the separation distance $r$ between their endpoints is the same. From the partition function for the chains, the free energy for the two chains combined is two times the standard stretch free energy, i.e. $\beta G_{\text{poly, ub}}(N, r) = 2 \times 3r^2/2Nb^2$. Note that this reference state free energy does not depend on the angle $\theta$ between the two chains.

The entropic free energy of the two chains is reduced when adding a bond at some combined monomer distance (i.e. loop length) $n_1 + n_2 = n$ from the shared permanent crosslink. The resulting expression for $G_{\text{poly, b}} (n; N, r, \theta)$ is derived in Supporting Information section \ref{sec:StretchedGaussianPolymers}. The entropic free energy loss to the two polymers upon reversible crosslinker binding at position $n_1 + n_2 = n$ is then
\begin{align}
	\Delta G_{\text{poly}} (n; N, r, \theta) = &G_{\text{poly, b}} (n; N, r, \theta) - G_{\text{poly, ub}}(N, r).
	\label{eqn:FreeEnergyBindFinal2}
\end{align}
Note that this expression now depends on the angle $\theta$ between the chains, as this affects how much the chains are perturbed when forming the bond at $(n_1, n_2)$.

\begin{figure*}
	\centering
	\subfigure[]{\includegraphics[width= 0.48\textwidth]{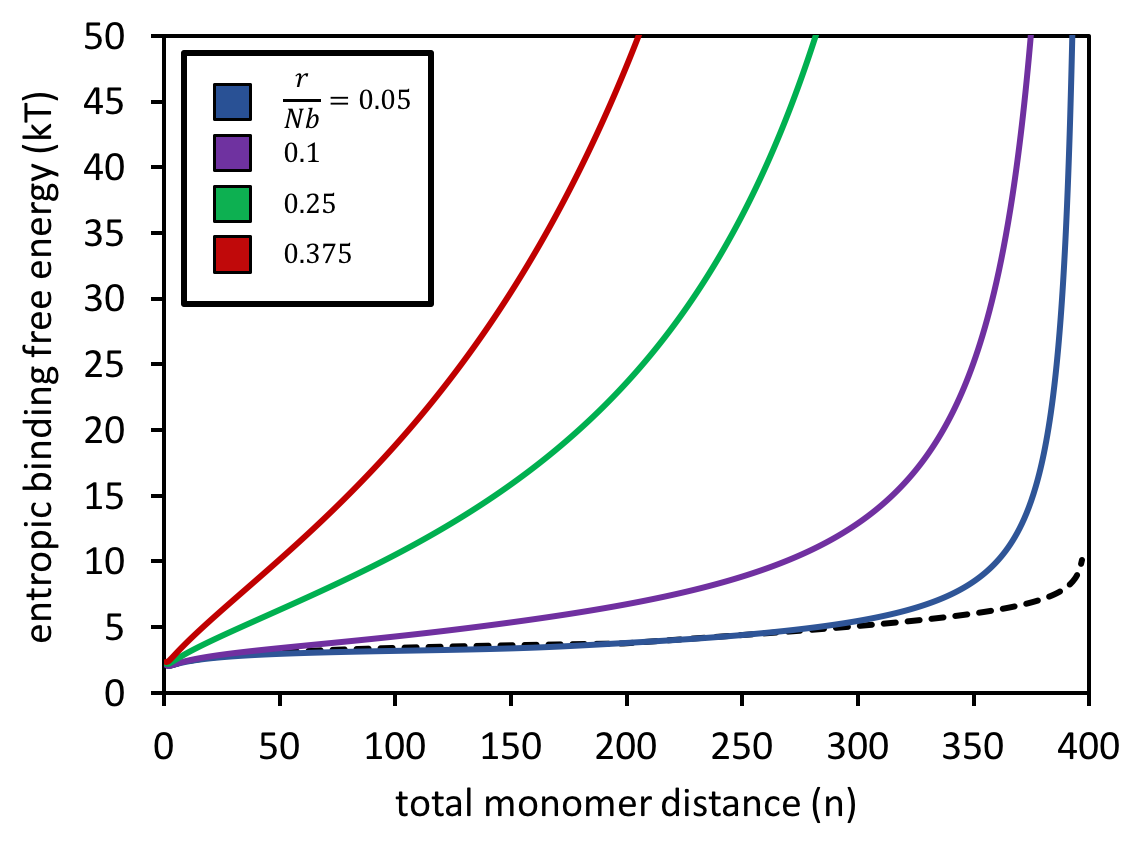}}
	\subfigure[]{\includegraphics[width= 0.48\textwidth]{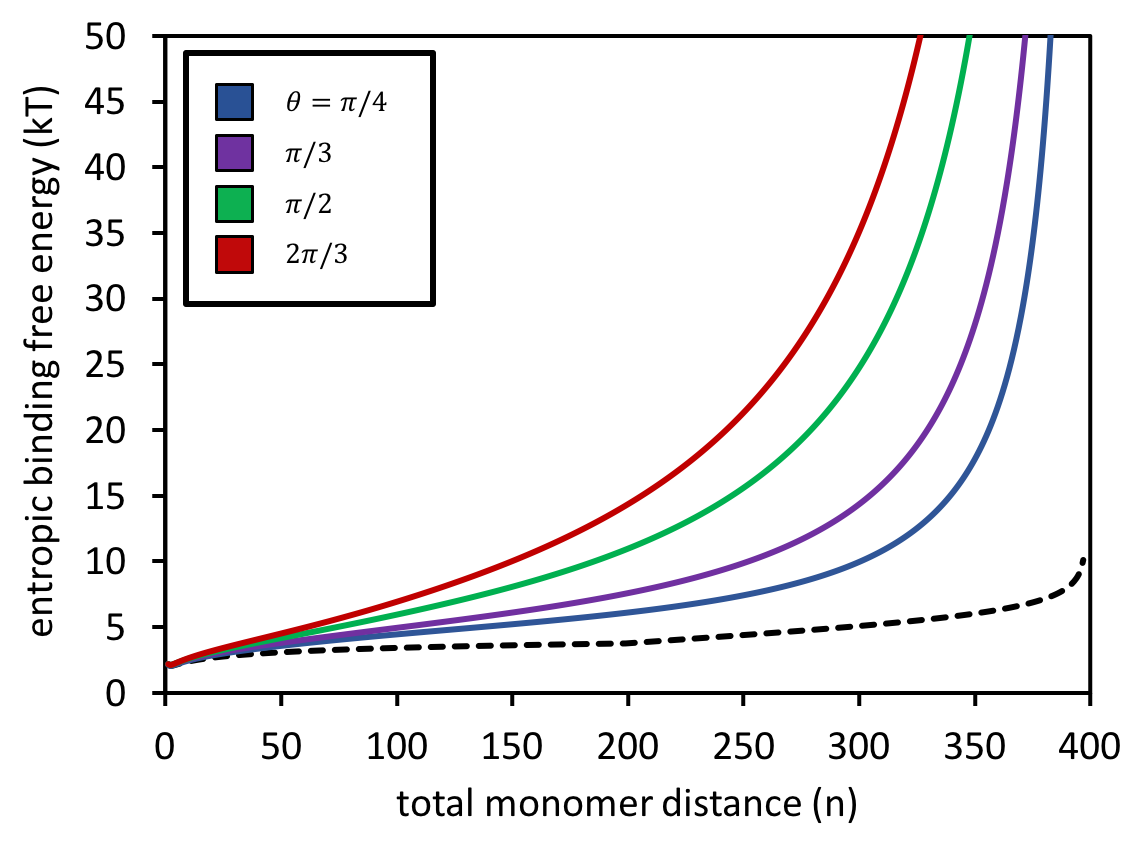}}
	\caption{Entropic binding free energy, $\beta \Delta G_{\text{poly}}(n; N, r, \theta)$, for forming a reversible crosslink at position $n_1 + n_2 = n$ along two polymers of $N = 200$ segments each, for different choices of chain stretch $r$ at constant angle $\theta$ (panel a), and for different $\theta$ at constant stretch $r$ (panel b).  In (a), the two chains are at an angle of $\theta = \pi/2$. In (b), the stretch ratio is set to $r/Nb = 0.15$.  In both panels, the black dashed curve is for when the two polymers have their ends untethered, i.e. Eq. \ref{eqn:FreeEnergyBindFinal}.}
	\label{fig:BindingEntropy}
\end{figure*}

Figure \ref{fig:BindingEntropy} presents calculations of the entropic free energy change, Eq. \ref{eqn:FreeEnergyBindFinal2}, for adding a bond at combined monomer distance $n$, for several choices of chain stretches $r$ and angles $\theta$. Results from the untethered chains (Eq. \ref{eqn:FreeEnergyBindFinal}) are also shown in each panel. Even for only moderately stretched polymers, such as the green dataset where the stretch ratio is $r/Nb = 0.25$, the entropy penalty grows very steeply with $n$. The penalty grows larger, to the order of tens of $kT$, as the distance between the chain ends grows (panel a) and also as the angle between them increases (panel b).

 For the polymer gel in contact with a non-depletable reservoir of reversible crosslinks at a fixed bulk concentration,  the full binding free energy for binding a reversible crosslink to a position $n_1 + n_2 = n$ is
\begin{align}
	\Delta G_{\text{bind}} (n; N, r, \theta) = &2 \Delta H_{\text{bind}} + \Delta G_{\text{cnf}} \nonumber \\ 
	&- \mu + \Delta G_{\text{poly}} (n; N, r, \theta)
	\label{eqn:FullFreeEnergy}
\end{align}
where $\Delta H_{\text{bind}}$ is the binding enthalpy for one polymer/reversible-linker bond, $\Delta G_{\text{cnf}}$ is any additional intra/inter-molecular configurational or interaction free energy cost for forming the transient connection, and $\mu$ is the chemical potential of the reversible linkers in the solvent phase of the gel. The latter is proportional to the natural logarithm of the concentration of reversible linkers in the solvent phase.

The \emph{reversible linker binding strength} and \emph{concentration} are two key parameters that dictate how much of an entropy cost the reversible crosslinks are able to ``pay''; that is, how far from the permanent crosslinks the reversible linkers are likely to bind. These two parameters can be manipulated to tune the overall equilibrium binding constant $K_{\text{eq}} = \exp{(-\Delta G_{\text{bind}}/kT)}$ of the linkers to the polymer backbone.

\begin{figure*}
	\centering
	\subfigure[]{\includegraphics[width= 0.90\textwidth]{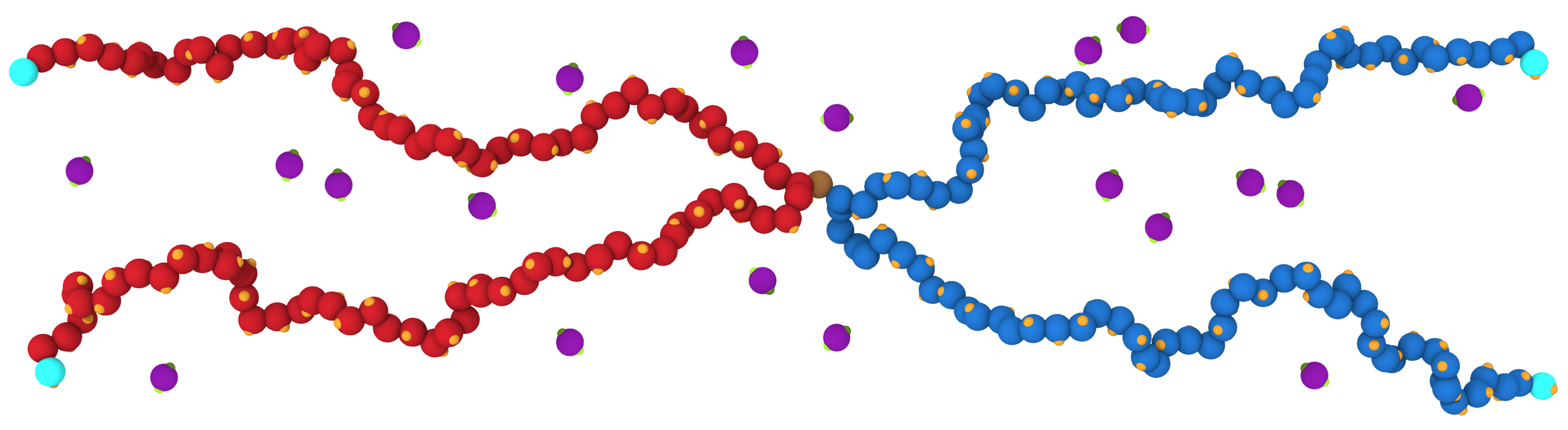}}
	\subfigure[]{\includegraphics[width= 0.90\textwidth]{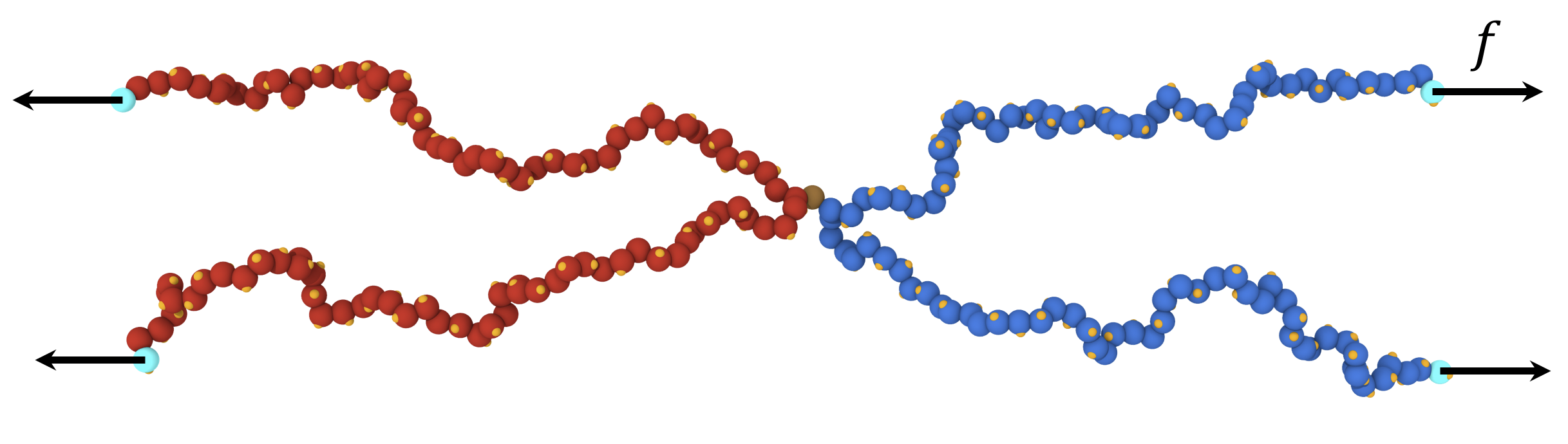}}
	\subfigure[]{\includegraphics[width= 0.90\textwidth]{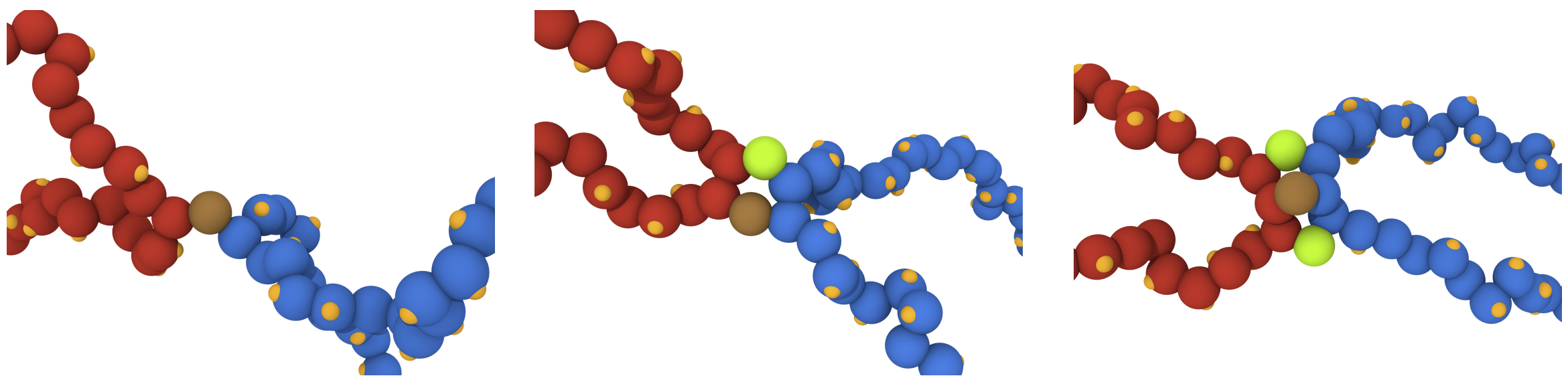}}
	\caption{(a) Snapshot of a molecular dynamics simulation of two polymers (blue beads, red beads) each of $100$ segments, permanently crosslinked at their midsection (brown bead). Reversible crosslink binding sites are shown in orange, and reversible crosslinking monomers are shown in purple, with green binding sites. (b) A force of strength $f$ can be applied to the light blue beads, with directions indicated by the black vectors. (c) Close-up around the permanent crosslink showing simulations with no helper linkers (left), one helper in green (middle), and two helpers (right). (Images generated with OVITO.\cite{Stukowski:2009ky})}
	\label{fig:MDSim}
\end{figure*}

A reversible linker binding strength and concentration that is sufficient to compensate for the entropy penalty of binding near a permanent crosslink, but not much further away, leads to selective clustering around the permanent links. This has two microscopic consequences:
\begin{itemize}
	\item{permanent crosslinks can share their local stress with adjacent bound transient crosslinks whilst the material is under strain;}
	\item{the polymer network topology is not substantially altered, as transient bonds far from permanent crosslinks have a low probability of formation and survival.}
\end{itemize}
At the macroscale these two effects should lead to a material which is tougher, yet responding elastically in a way that is very similar to that without the reversible crosslinks. The remaining discussion is dedicated to demonstrating this connection with molecular simulation and numerical modelling. On the other hand, very negative and favourable $\beta \Delta H_{\text{bind}}$ or a very high concentration of reversible linkers will lead to non-selective binding, as the entropy contribution becomes negligible regardless of the binding position $n$. Reversible crosslink binding will then be non-selective, leading the network topology to be altered, and the macroscopic elasticity to be different than the native material without the reversible crosslinks.

We have thus far cast our theory in terms of thermodynamic entropy. Similar qualitative conclusions could be obtained by arguing that the reversible crosslinks cluster around the permanent crosslinks on the grounds of a \emph{structural proximity} effect,  particularly if the gel is quite dilute . The argument would be that near a permanent crosslink, two polymer strands are by necessity already in close proximity. Therefore, there is a higher likelihood of adding a connecting reversible crosslink in that vicinity compared to elsewhere  in a dilute gel .

While conceptually sound, the line of thinking based on proximity effects does \emph{not} yield an obvious mathematical framework for quantifying the recruitment effect relative to the other key thermodynamic factors involved in reversible crosslinker binding, namely the binding strength, and reversible linker concentration. Casting our argument fundamentally in terms of entropy, on the other hand, gives a clear and more general quantitative handle (e.g. Eq. \ref{eqn:FullFreeEnergy}) for assessing the balance between these thermodynamic factors, so we can justify which choices of microscopic design give rise to the optimal sought-after macroscopic behaviour.

\section{Clustering \& load sharing around permanent crosslinks}

In this section, we use molecular dynamics to examine reversible crosslink recruitment around a permanent crosslink, and to what extent adjacent bound reversible links share stress at a permanent crosslink.

Illustrated in Figure \ref{fig:MDSim}(a), the model consists of two polymer chains, each of 100 segments, connected together at their midsection (segment 50). The chains are placed in a simulation box, along with a given number of bivalent reversible crosslinkers. Details of the model are given in Supporting Information section \ref{sec:MDDetails}; what follows is a brief summary of the ingredients. The polymer segments are connected together by strong harmonic bonds. Non-bonded monomers interact via a repulsive inverse power law  potential $V(r)\propto r^{-12}$ (with $r$ the intermonomer distance). Each monomer has attached to it a binding site. The reversible crosslinkers each consist of a single bead, with two binding sites attached, held at an angle of $\pi$ relative to each other by a strong three-body angle potential.  Binding sites on a reversible crosslink interact with binding sites on the polymers via an attractive Gaussian potential. Calculations are carried out using the HOOMD-blue molecular dynamics software package.\cite{Anderson:2008bt, Glaser:2015cu}

At present, we are interested only in the statistics of reversible crosslink binding between the \emph{different} polymers (red and blue) in Figure \ref{fig:MDSim}(a). In order to study just the physics of intermolecular reversible crosslink binding, and to prevent binding between segments on the \emph{same} chain, the model distinguishes between binding sites on the red and blue polymer. The two binding units on a given reversible linker are also distinct, so that one can bind only to binding sites on the red polymer, while the other only to those on the blue polymer.

\begin{figure}[h]
	\centering
	\subfigure[]{\includegraphics[width= 0.48\textwidth]{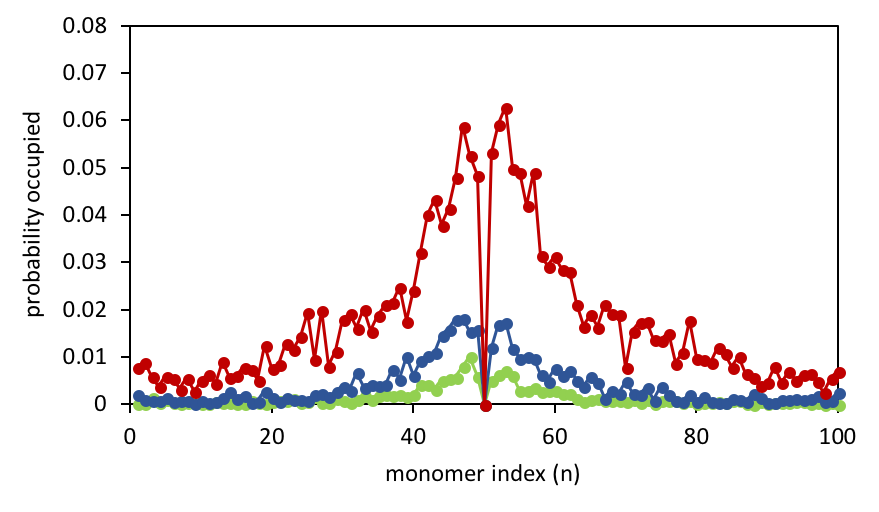}}
	\subfigure[]{\includegraphics[width= 0.48\textwidth]{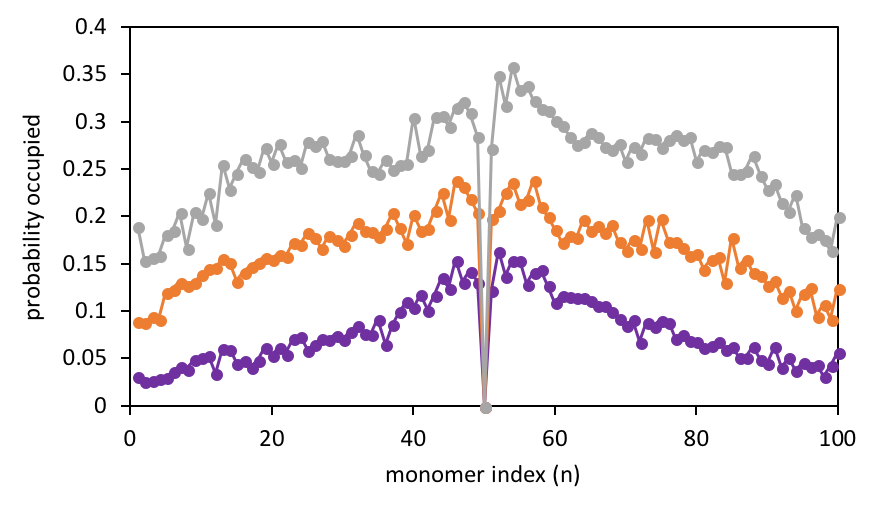}}
	\caption{Simulation-averaged probability that a doubly bound reversible crosslink is attached to monomer $n$. Curves coloured green, blue, red in panel (a), and purple, orange, grey in panel (b), are for increasing reversible crosslink binding strength  $\beta \epsilon_{\text{bind,eff}} = -12.9$, $-13.4$, $-13.9$, $-14.4$, $-14.9$, and $-15.5$. Distributions are normalised so that a value of unity at a given monomer index $n$ means that a doubly bound reversible crosslink is \emph{always} bound to that monomer throughout the duration of the simulation, while a value of $0$ means that there is \emph{never} a reversible crosslink bound to that monomer. }
	\label{fig:SimBindingProb}
\end{figure}

The model is employed to examine the equilibrium statistics of reversible crosslinker binding along the polymer chains. In Figure \ref{fig:SimBindingProb}, we record the simulation-average probability that segment $j$ (from $1$ to $100$) is bound to a doubly bound reversible crosslink, for six different choices of reversible crosslink binding strength. At low and intermediate binding strength, we see very clear preference for binding near the permanent crosslink $j = 50$. The reversible crosslinks form a ``zipper'' domain around the permanent crosslink, resembling what is observed in earlier molecular simulation studies.\cite{Kindt:2005gb} (Indeed, here the permanent crosslink is acting like a nucleation site for the zipper domain.) At larger binding strength, the reversible linkers bind more randomly across the polymers.

\begin{figure}
	\centering
	\includegraphics[width= 0.48\textwidth]{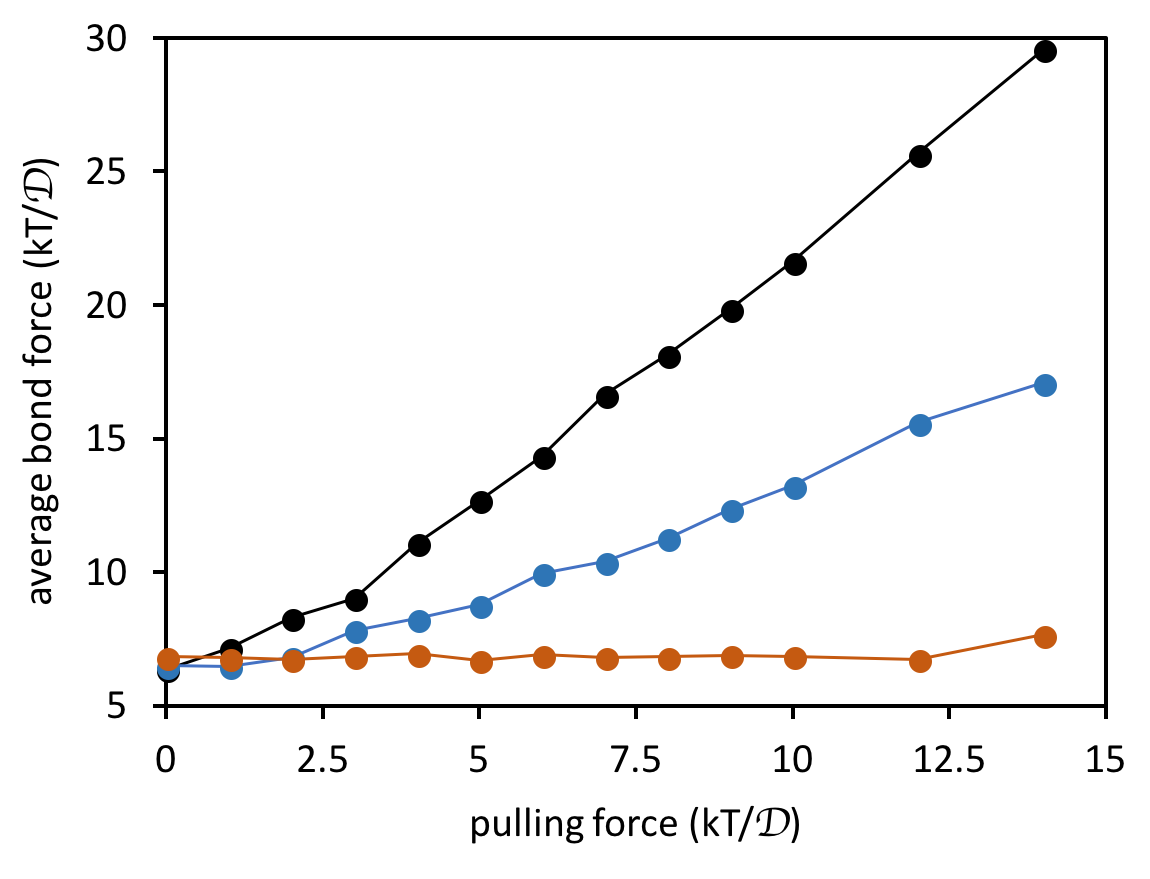}
	\caption{Average force on the permanent crosslink as a function of pulling force $f$ on the four chain ends. Black curve is for system with only the permanent crosslink, blue includes one adjacent helper, and orange includes two adjacent helpers. See Figure \ref{fig:MDSim}(b) for simulation snapshots of the three scenarios.}
	\label{fig:SimLocalStrength}
\end{figure}

We now turn to examining how bound linkers adjacent to the permanent crosslink locally share stress. To do so, the reversible crosslinking units are removed from the system, and one or more linkers are permanently affixed to the polymer segment(s) adjacent to the permanent crosslink as shown in Figure \ref{fig:MDSim}(c). These ``helper'' linkers remain bound to their respective polymer segments for the duration of the simulation. The four ends of the polymers are then pulled with a constant force $f$ in opposite directions throughout the duration of the simulation. As depicted in Figure \ref{fig:MDSim}(b), the ends of polymer A are pulled along $+x$, and those of polymer B along $-x$; they are otherwise free to fluctuate in $(x, y, z)$. This results in an extensional force along the permanent crosslink.

This simulation setup corresponds to a realistic physical scenario in which the bond exchange \emph{kinetics} of the reversible crosslinkers are \emph{slow}; that is, the average lifetime for a reversible crosslink to remain attached to its two partners is far longer than the (microscopic) simulation timescale. Indeed, in experiment \cite{Kean:2014hh} the reversible crosslink bonding half-life spans the order of milliseconds to seconds, clearly far longer than the molecular timescale being examined in our molecular dynamics simulation. As such, fixing the helper linkers in position over the course of our simulation is actually most representative of the experimental regime of interest for mechanical testing at the microscale.

\begin{figure}
	\centering
	\subfigure[]{\includegraphics[width= 0.45\textwidth]{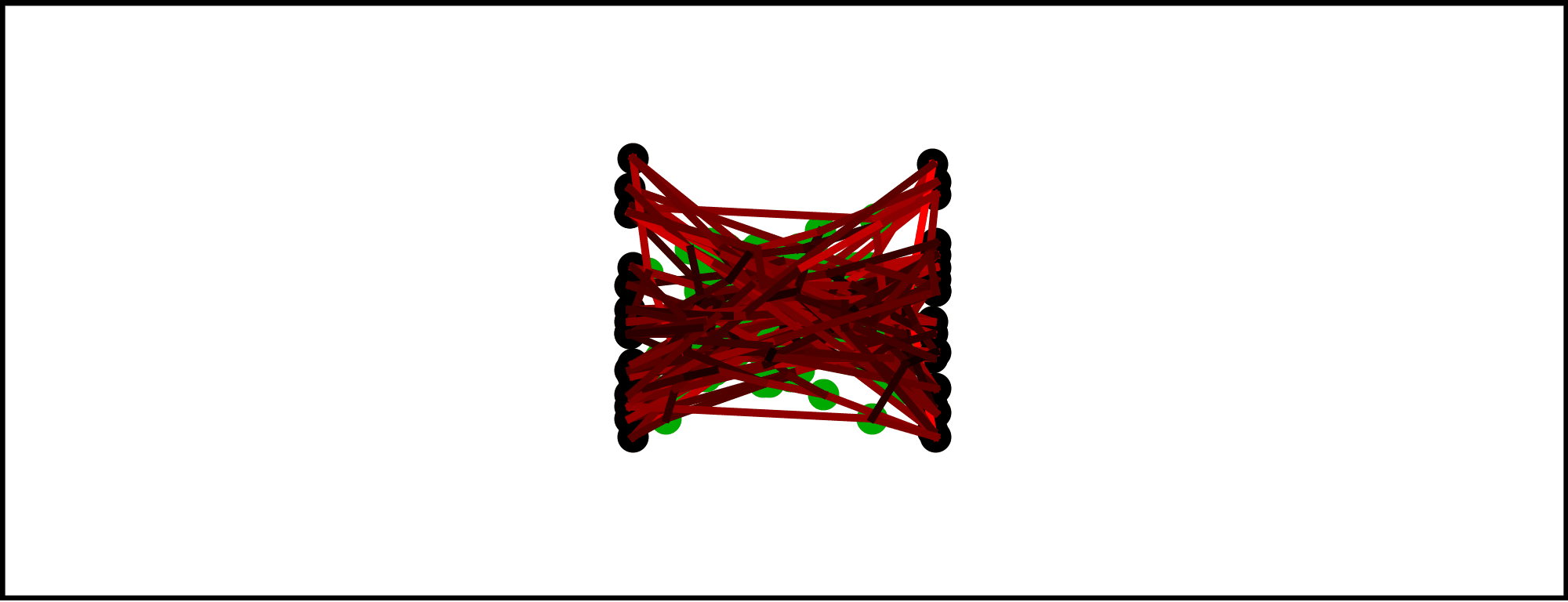}}
	\subfigure[]{\includegraphics[width= 0.45\textwidth]{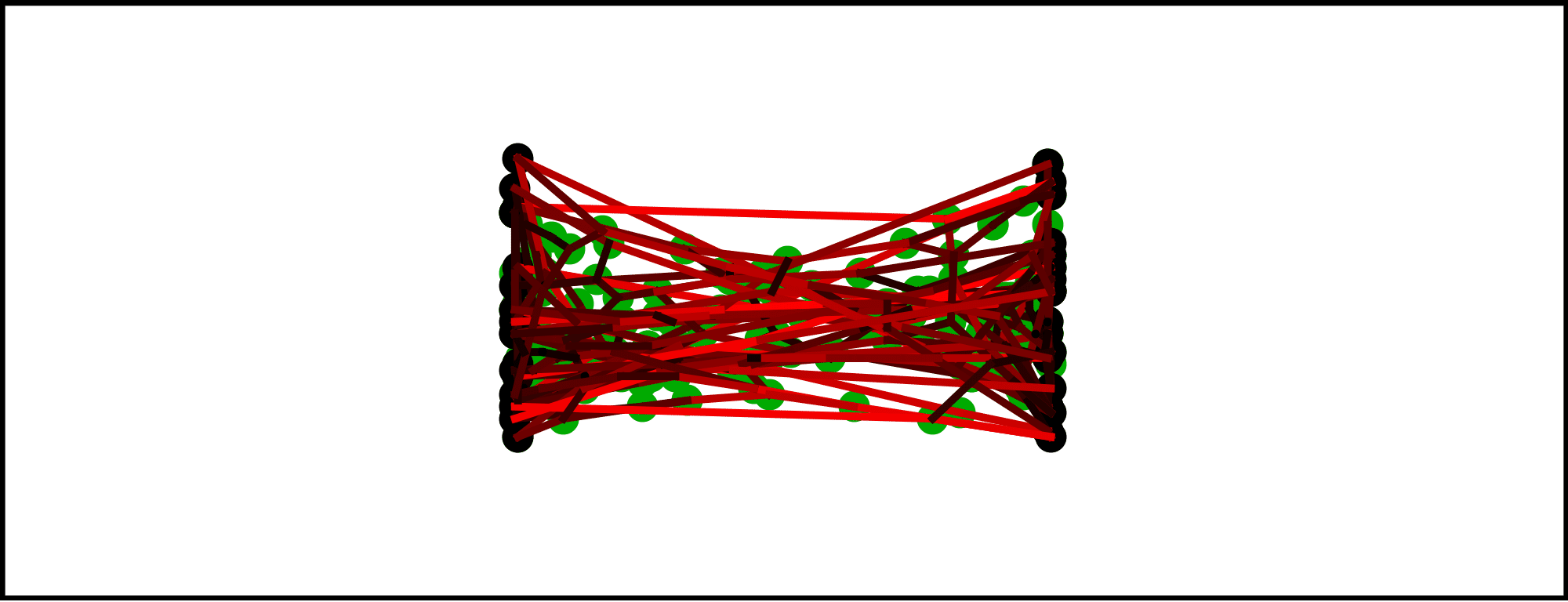}}
	\subfigure[]{\includegraphics[width= 0.45\textwidth]{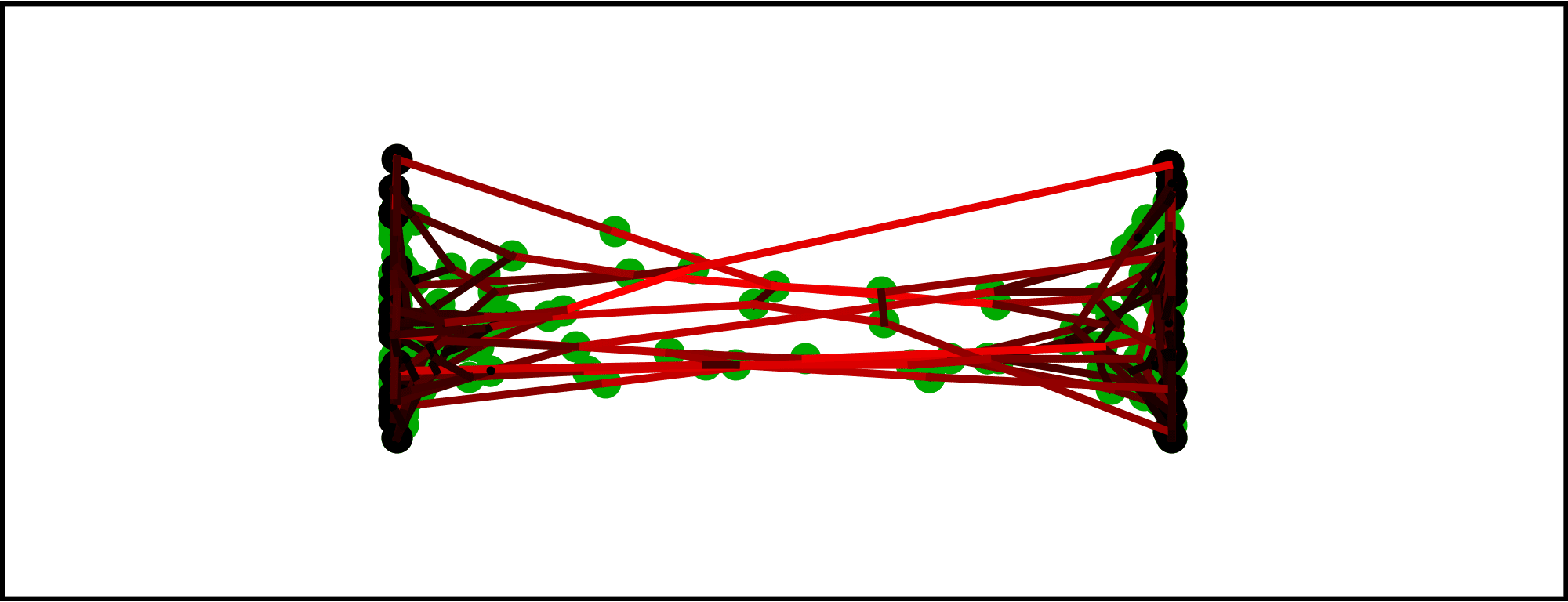}}
	\subfigure[]{\includegraphics[width= 0.45\textwidth]{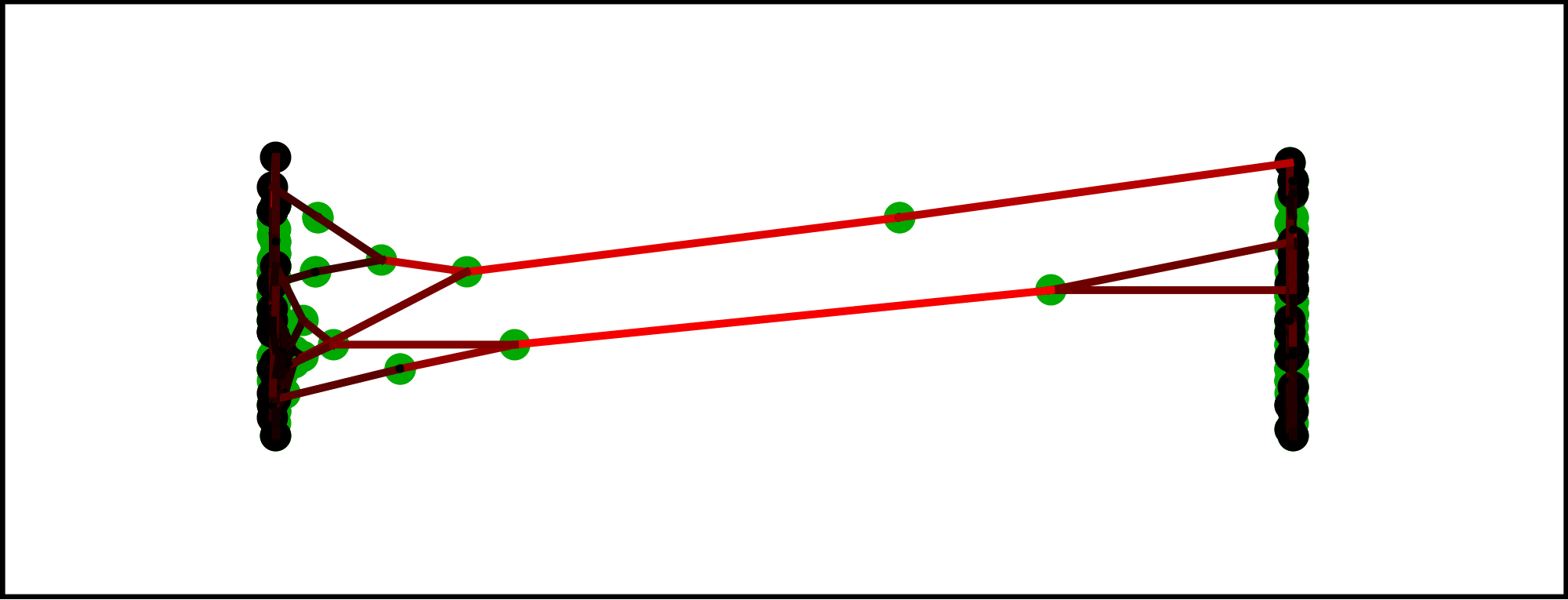}}
	\caption{Snapshots of a network strain experiment in the lattice SCFT model, going from low strain (a) to high strain (d). Crosslinks (nodes) are shown as green dots, localised to their most probable position as given by the SCFT calculation, and polymer chains (bridges) are represented by lines. The shading of the line indicates that the polymer is experiencing high (red) or low (black) tension following Eq. \ref{eqn:BridgeForce}. The network is suspended between two fictitious plates by the solid black points. Polymer bridges disconnect from one of their two attachment nodes when the force on the bridge exceeds $F^* = 1.25 kT / b$, in this case.}
	\label{fig:NetworkSnapshots}
\end{figure}

Importantly, as will be discussed in Section \ref{sec:Kinetics}, reversible bonding exchange \emph{kinetics} are not necessarily coupled to the \emph{thermodynamic} linker binding strength. Slow kinetics, with long reversible linker binding lifetimes, can result from large activation barriers (entropic or enthalpic in origin) for binding or unbinding. These barriers can be present and independently tuned even when the linkers have a ``weak'' thermodynamic binding free energy, as is necessary for entropy to guide recruitment of the linkers around permanent crosslinks. We will argue that this is actually the optimal recipe for toughness-enhancing reversible crosslinks in Section \ref{sec:Kinetics}. 

The  average force along the permanent crosslink  is obtained from simulation by monitoring the average lengths $\langle l \rangle$ of its two bonds, relative to the preferred bond length $l_0$. The bonds follow the harmonic force law $\langle F \rangle = k (\langle l \rangle - l_0)$, where the spring constant $k = 1000 \ kT/\mathcal{D}^2$ and $\mathcal{D}$ is the simulation length unit. From $\langle l \rangle$ in simulation, $\langle F \rangle$ may be calculated.

Figure \ref{fig:SimLocalStrength} reports the average force $\langle F \rangle$ along the permanent crosslink, as a function of pulling force $f$, for: the two polymers connected only by the permanent crosslink; with one helper; and with two helpers. The addition of the helpers indeed reduces the average tension along the permanent crosslink bonds. If we suppose that the probability of permanent crosslink failure grows exponentially with its average force according to the Bell model \cite{Bell:1978hj, Evans:1991hu},
\begin{equation}
	P(\text{fail}) = A \exp{\left(B \langle F \rangle / kT\right)},
\end{equation}
where $A$ and $B$ are constants, then the presence of the helper linkers substantially reduces the probability of bond breakage. Indeed, adding more helpers leads to a mutually lower probability that any one of the linkers or the permanent crosslink will fail at a given pulling force.\cite{Seifert:2000uk}

\begin{figure*}
	\centering
	\subfigure[]{\includegraphics[width= 0.49\textwidth]{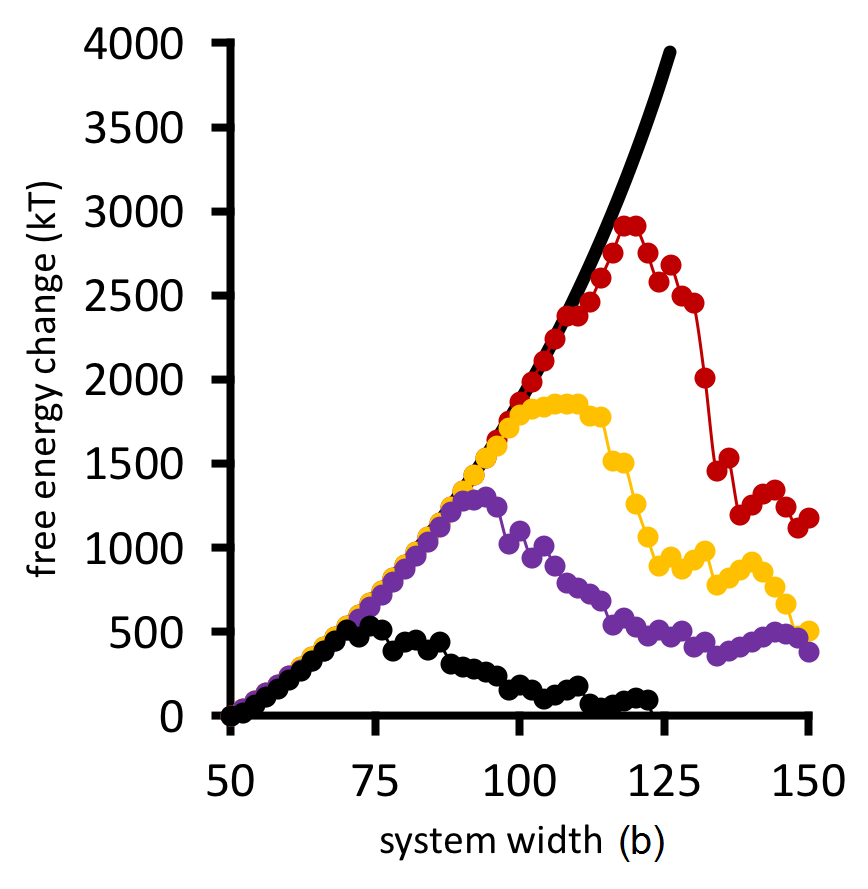}}
	\subfigure[]{\includegraphics[width= 0.49\textwidth]{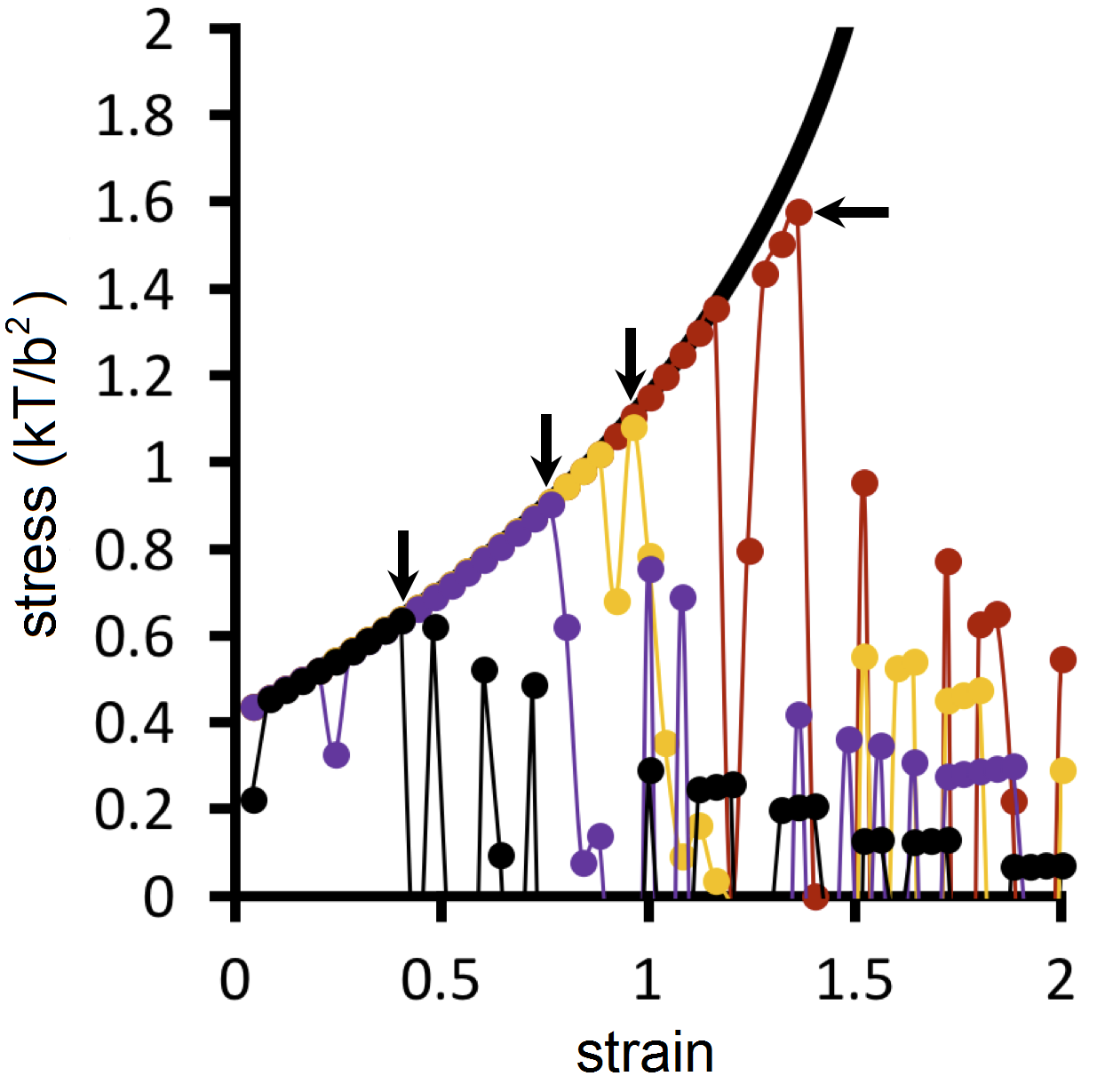}}
	\caption{Network free energy (a) and stress (b) as a function of strain (horizontal system size), for different choices of $F^* / F^*_\circ = 1.0, 1.2, 1.3, 1.4$ (points connected by lines in black, purple, yellow, and red respectively). Solid black line is for when bridges are prohibited from breaking during strain. Arrows in (b) indicate the rupture point of the material. Value of $F^*_\circ = 1.25 kT / b$.}
	\label{fig:MPResults}
\end{figure*}

\begin{figure}
	\centering
	\includegraphics[width= 0.48\textwidth]{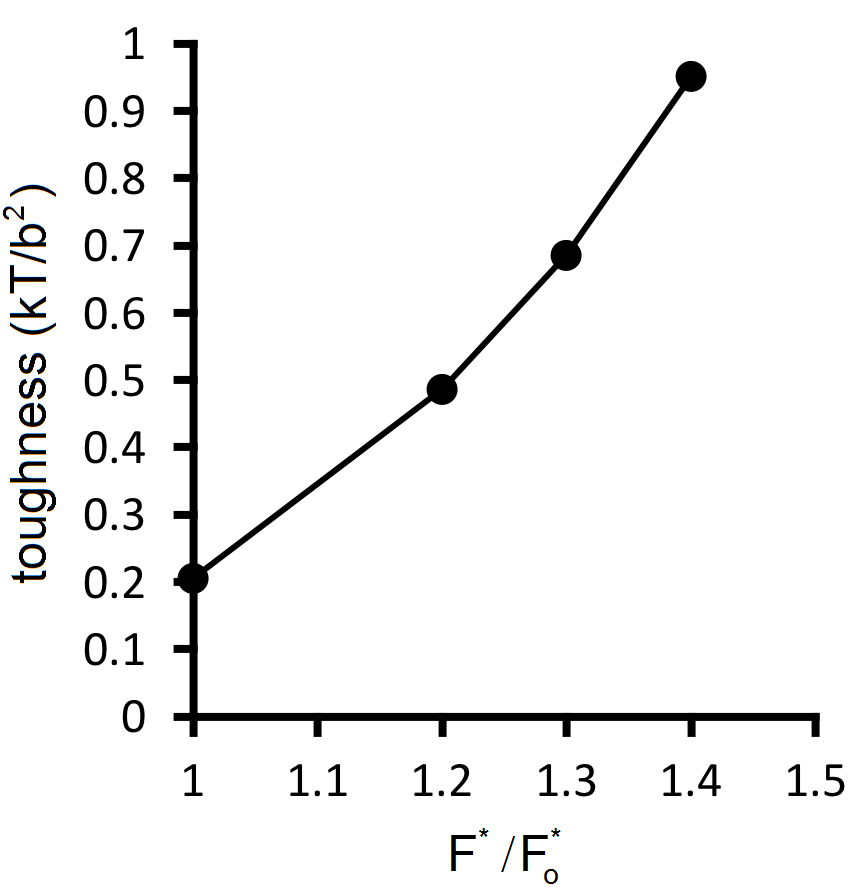}
	\caption{Stress-strain toughness of the network in SCFT modelling, as a function of crosslink strength $F^*$ relative to $F^*_\circ = 1.25 kT / b$.}
	\label{fig:Toughness}
\end{figure}

\section{Amplified toughness enhancement at the network scale}

Recruitment of reversible crosslinks around permanent crosslinks implies that the latter are locally reinforced, so that they can bear more force before breaking. In this section, we examine the consequences of locally reinforced crosslinks on the global toughness of a polymer network as it is strained. To be able to apply strain at a timescale that is much longer than the relaxation time of polymer strands in the network, a coarse-grained self-consistent field theory (SCFT) model for polymer networks that we previously developed is utilised.\cite{Tito:2017db}

The model uses lattice SCFT \cite{Scheutjens:1979ib} to converge on the equilibrium distribution of crosslinks and polymer chain configurations in a microscopically sized volume element of the material, given a fixed local network topology (polymer chain connectivity) and boundary conditions.  The network is represented with finitely extensible phantom (i.e. non-interacting non-self-avoiding) lattice polymers, and does not account for chain entanglements.  Each polymer is composed of a given number of segments of width $b$, equal to the width of a single lattice site. 

The model allows for easy extraction of the free energy of the network, as a function of strain applied to the boundaries of the system. For simplicity, we perform the calculations in two dimensions. Even though the entropic recruitment of reversible crosslinks near permanent ones is a three-dimensional phenomenon, we expect the effect of crosslink strength on global toughness to be similar in two and three dimensions.

The calculations carried out here consist of defining a square polymer network with an initial width and height within the lattice SCFT model, and then straining along its horizontal axis via a series of strain steps. On each strain step, the SCFT model is used to approximate the equilibrium spatial distribution of polymers and crosslinks for the current system dimensions, obtaining the network free energy for that strain. On the next strain step, the system width is increased by a desired amount, and the process repeated.  Thus, the SCFT model represents the limiting scenario of a \emph{quasi-equilibrium} strain experiment on the polymer network.

In order to examine how local toughness translates into global toughness, we must microscopically capture when and where connections break within the network as it is strained. Our model from Ref. \citenum{Tito:2017db} is adapted by adding in the possibility for polymer chains (``bridges'' in the model) to irreversibly disconnect from one of its two attachment points (``nodes'') as the network boundaries are strained. The bridge is instantaneously and irreversibly cut when its tension exceeds $F^*$, a tunable threshold tension parameter.

The parameter $F^*$ physically represents the effective strength of the crosslinks in the network. It is into this parameter that the physics of reversible crosslinking outlined in the previous two sections can be invested in a coarse-grained way.

We begin by choosing a baseline value of $F^* = F^*_\circ$ representing the strength of the permanent crosslinks alone. Increasing the value of $F^*$ from this baseline $F^*_\circ$ represents the idea that reversible crosslinks have attached next to these permanent crosslinks, locally enhancing the strength of the connections. In the following calculations, we assume that $F^*$ is the same for every connection in the system. This is for simplicity; in reality, the $F^*$ of each connection is different, depending on the number of reversible crosslinks that have recruited around the connection.  Any mechanical contributions that would arise from reversible crosslinks binding far from permanent crosslinks, thereby altering the topology of the network, are neglected. Thus, the assumption in this model is that the reversible crosslinks have a weak-binding $K_{eq}$, and are entropically driven to bind only near permanent crosslinks as discussed in Section II. 

For the following calculations, a square network initially of $50 \times 50$ lattice units in size is defined with $283$ bridges, and $190$ nodes. Details of the network configuration are given in Supporting Information section \ref{sec:SCFTDetails}.  The network topology and number of crosslinks is the same for all calculations, while only the crosslink strength parameter $F^*$ is varied. 

Figure \ref{fig:NetworkSnapshots} shows a schematic of the polymer network used for calculations. The figure shows the polymer network in its initial state, and at different subsequent strain steps. The polymer bridges (lines) are shaded to indicate whether the force in the bridge is low (black) or high (red). As the network is strained,  the tensions in the bridges grow depending on their average end-to-end distance $L_i$. Provided we operate within a regime where bridges are not stretched very near to their finite end-to-end extent (accomplished by choosing an appropriately moderate value of $F^*$), then the force along a bridge can be approximated by the ideal chain model in two dimensions:
\begin{equation}
	F_i(L_i) =  \frac{2 L_i k T}{N_i b^2},
	\label{eqn:BridgeForce}
\end{equation}
where $N_i$ is the number of segments in the bridge, and $b$ is the monomer width (taken to be unity). A bridge breaks away from one of its two permanent crosslinks when its tension exceeds $F^*$. In this system shown, $F^* = 1.25 kT / b$; this is hencforth defined as the ``baseline'' system, with $F^*_\circ$ equal to this $F^*$.

Figure \ref{fig:MPResults}(a) plots the free energy of the network as a function of system width,  while Figure \ref{fig:MPResults}(b) shows the stress as a function of strain. The strain is defined as the change in system width, divided by its initial width ($50b$). The stress, in units of $kT / b^2$, is defined as the finite difference derivative of the free energy in Figure \ref{fig:MPResults}(a) divided by the fixed height of the system ($50b$). 

Numerical results for the baseline system are plotted in black points in Figure \ref{fig:MPResults}. The remaining datasets show calculations for the same network, but with larger choices of $F^*$ relative to the baseline dataset. The free energy in Figure \ref{fig:MPResults}(a) initially increases with strain; if none of the bridges are allowed to break, then the free energy follows the solid black line. The increase of the free energy with strain is sharper than quadratic in this network, because the model captures the finite extensibility of the polymers \cite{Tito:2017db}. The restoring force as a function of strain for the network therefore increases faster than linearly in Figure \ref{fig:MPResults}(b). 

Upon continued strain in Figure \ref{fig:MPResults}, the network reaches a critical rupture point, defined as the strain where the stress reaches a maximum (black arrows in Figure \ref{fig:MPResults}(b)). At or just before the critical strain, the free energy of the system begins to fluctuate and decrease in a jagged fashion. Each downward plunge in the free energy corresponds to breakage of a bridge/node connection.

The toughness of the network is defined as the integral of the stress-strain curve in Figure \ref{fig:MPResults}(b) prior to rupture. These results are given in Figure \ref{fig:Toughness} for the four choices of $F^*$ studied here. The toughness of the network grows faster than linearly with $F^*/F^*_\circ$.

To build on this observation, compare the strain results for different $F^*$ in Figure \ref{fig:MPResults}. Indeed we find that the stress the network can bear before rupturing relative to the baseline system grows \emph{faster} than the factor $F^* /  F^*_\circ$ of local crosslink reinforcement. For example, the red dataset in Figure \ref{fig:MPResults} uses $F^* / F^*_\circ = 1.4$, i.e. the crosslinks can locally bear $1.4$ times more tension before breaking, compared to the baseline system. However, this local enhancement allows the network to globally bear $\approx 2.5$ times more stress relative to the baseline system before rupturing (Figure \ref{fig:MPResults}b). For the yellow dataset, $F^* / F^*_\circ = 1.3$, while that network can bear $\approx 1.7$ times more stress compared to the baseline system.

These calculations suggest that reinforced crosslinks at the molecular scale leads to an even larger factor of toughening for the whole network. This is likely a cooperative effect, whereby reducing the probability for breaking a single isolated connection translates, in the network context, into an even lower chance of failure of a connection that is bearing stress shared by other nearby connections \cite{Evans:1991hu, Seifert:2000uk, Vaca:2015cn}. Thus, reversible crosslinks that locally reinforce the permanent crosslinks of a polymer network even by a small margin can lead the network to be tougher by a larger factor, at the macroscopic scale. 

\begin{figure}
	\centering
	\includegraphics[width= 0.48\textwidth]{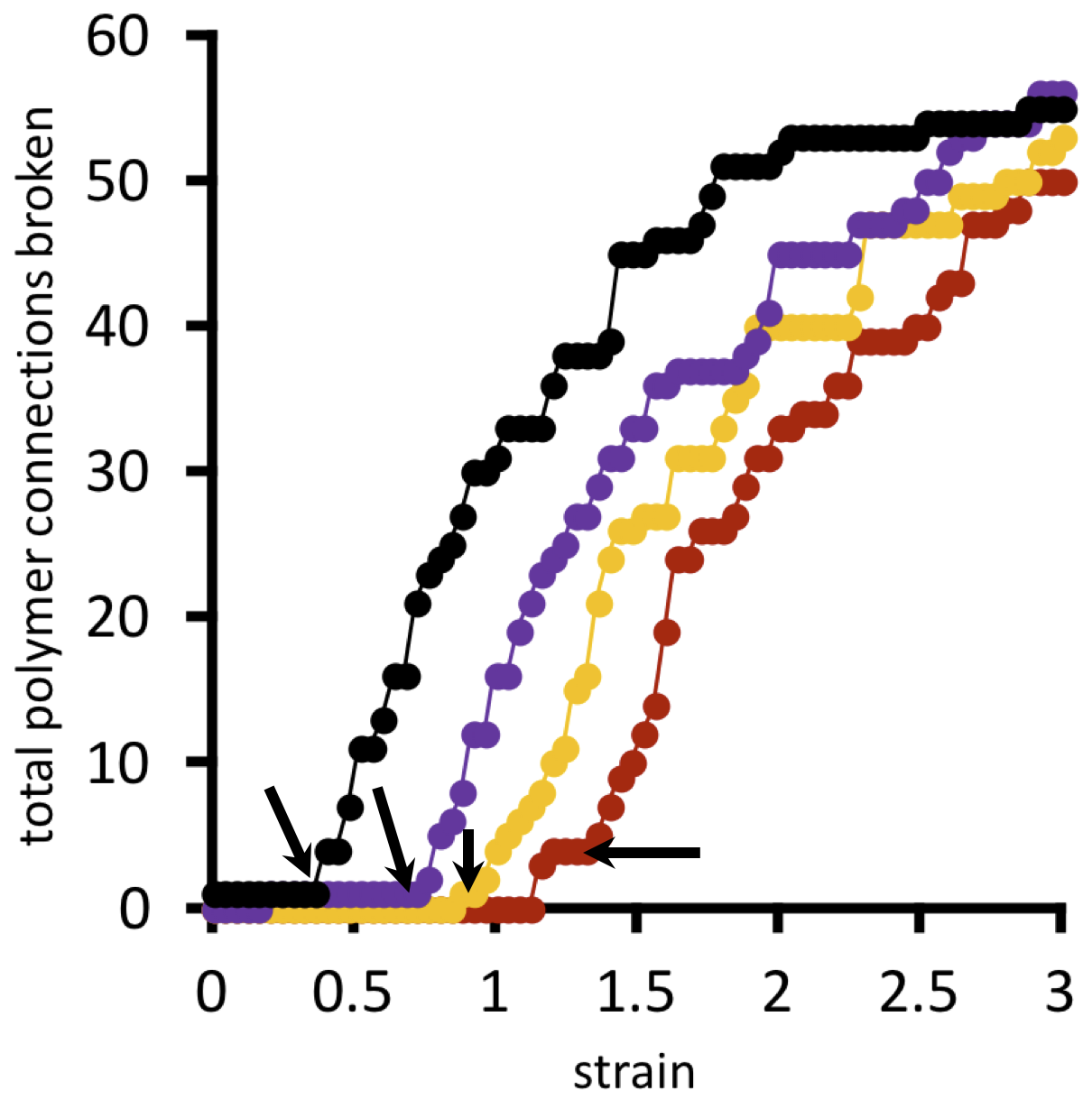}
	\caption{Total number of polymer bridges with broken connections as a function of strain, for different choices of $F^* / F^*_\circ = 1.0, 1.2, 1.3, 1.4$ (points connected by lines in black, purple, yellow, and red respectively). Arrows indicate the rupture point of the material. Value of $F^*_\circ = 1.25 kT / b$.}
	\label{fig:MPResults2}
\end{figure}

A polymer bridge breaks when its force, calculated by Eq. \ref{eqn:BridgeForce}, exceeds the given choice of the breakage force parameter $F^*$.  Figure \ref{fig:MPResults2} presents the cumulative number of polymer bridge connections broken in the network as a function of horizontal strain. This provides an assessment for the mechanism by which the network fails when approaching and surpassing its rupture point (indicated by black arrows in the figure). Before the rupture point, polymer bridge failure is rare, and the network remains largely intact. Beyond the rupture point, polymer bridges rapidly (and irreversibly) break away from their permanent crosslinks.  In this quasi-equilibrium model, where strain is imagined to happen at an infinitely slow rate, the number of bridge failures per unit strain generally follows the same curve for all choices of crosslink strength $F^*$ considered; the choice of $F^*$ only shifts the curves horizontally, i.e. the strain at which rupture initiates. Therefore, locally strengthening the crosslinks in the network (i.e. by larger $F^*$) serves only to delay the onset of rupture, while not having an influence on the rate of polymer bridge rupture with respect to strain thereafter. 

The topology of the network, among other factors, dictates to what extent local stress is transferred through the network as it ruptures, which will in turn temper the influence of local crosslink reinforcement on global toughness. A recent simulation study by Nabavi et al.\cite{Nabavi:2015ki} considered the work required to stretch a polymer chain to full extent, when reversibly associating monomers are embedded into the polymer chain in a particular sequence. Altering the reversibly bonding monomer sequence, so that the polymer exhibits a different self-associated conformation in the unstretched state, can shift the work required to rupture the links and stretch the polymer by a factor of two to three. This order of magnitude is consistent with the degree of toughening amplification observed for our network in Figure \ref{fig:NetworkSnapshots}, though a systematic examination of toughness in different network topologies should be carried out to gain better insight into this effect.

\section{Experimental guidelines \& kinetic considerations}
\label{sec:Kinetics}

We have developed a microscopic theory for freely diffusing reversible crosslinks in a permanently crosslinked polymer gel. The theory enables us to define clear, albeit qualitative, guidelines for optimal reversible crosslink design in experiment.

When the equilibrium constant for reversible crosslink binding to the polymer chains is small, then entropy dictates the ensemble of binding configurations at equilibrium. Entropy favours clustering of the reversible links around permanent crosslinks. This leads to local load sharing of stress around the permanent crosslinks during strain, so that the material has a globally lower probability of rupture at higher stress relative to the native material. This translates into higher material toughness. Moreover, the microscopic topology of the polymer network is not affected by the reversible crosslinks, and so the elasticity of the material is left mostly unaltered compared to the native material.

On the other hand, if the reversible link equilibrium binding constant is large, then the entropic bias becomes negligible. The reversible links are then expected to bind in random locations to the polymer network scaffold. The altered network topology at the microscale results in a different macroscopic elasticity compared to the native material, and the permanent crosslinks are not selectively reinforced at the microscale by reversible links. In this regime, material toughness is not enhanced. 

This picture is consistent with the experimental results in Kean et al.\cite{Kean:2014hh} In their study, relatively weak-binding reversible crosslinks lead to the greatest toughness enhancement, with the smallest change in elasticity. This is in contrast to their strong-binding linker, which causes the material to become stiffer, and more brittle.

Our theory is also in agreement with the recent experimental work by Mayumi et al.\cite{Mayumi:2013gd, Mayumi:2016kr} This study examines the behaviour of a permanently crosslinked gel, with a fixed concentration of reversible crosslinks, at different strain rates. In fracture tests of notched samples, the gel exhibits the least change in low-strain modulus and greatest enhancement of intrinsic toughness, compared to the gel without reversible crosslinks, in the limit of slow loading.  The sample is also far more extensible before fracture in this case. 

In the context of our theory, the slow-strain limit allows the reversible crosslinks sufficient time to equilibrate---to sample a broad ensemble of microscopic configurations---in which reversible crosslinks are localised around the permanent crosslinks as the polymer strands are stretched. While the reversible linkers will be dynamically moving and exchanging binding partners in a slow-strain limit experiment, the macroscopic stress-strain behaviour of the material will be dominated by the thermodynamic ensemble average of reversible crosslink binding. If, on the other hand, the material is strained more rapidly, then the timescale of strain will begin to approach and exceed the timescale of reversible crosslink rearrangement. The material will be more strongly governed by reversible crosslink configurations that are kinetically locked on the timescale of the experiment (for example, see Ref. \citenum{Yount:2005cv}).

Reconfigurability below the strain timescale is where reversible crosslinks present their greatest utility in the present case. As a juxtaposed example, consider instead adding new permanent crosslinkers to the native polymer gel. New connections between the polymer scaffold will be formed, but the locations of the new bonds are largely dictated by first-passage kinetics. Once a permanent crosslink finds a partner, it is stuck. Entropy instead acts at the ensemble level, when added crosslinks are allowed to try and re-try many different binding options within the fluctuating polymer network. This provides the opportunity to tune the linker binding strength and concentration, so as to bring them into the entropically governed regime where linkers localise around the permanent crosslinks.

We have thus far argued that a small equilibrium constant for reversible linker binding is necessary for entropic recruitment around permanent crosslinks. However, the reversible crosslinks must also have sufficiently strong and long-lived bonds so that they are useful in sharing local stress near a permanent crosslink. Indeed, in Kean et al., half-lives of binding for the reversible crosslink structures considered are on the order of milliseconds to seconds.\cite{Kean:2014hh} In microscopic terms, these are very ``strong bonds''. We now decipher how strong-binding linkers may still be within a ``weak-binding'' regime.

The equilibrium binding constant, related to the binding free energy by the standard relation $\Delta G_{\text{bind}} = -RT \ln{K_{\text{eq}}}$ (where $R = 8.314 \ J \ \text{mol}^{-1} K^{-1}$ is the ideal gas constant), is in turn determined by the ratio of the rate $k_{\text{on}}$ of reversible linker binding to the rate $k_{\text{off}}$ of unbinding:
\begin{equation}
	K_{\text{eq}} \propto \frac{k_{\text{on}}}{k_{\text{off}}}.
	\label{eqn:KeqExpt}
\end{equation}
The rate of unbinding $k_{\text{off}}$ can be directly tuned by the binding \emph{enthalpy} $\Delta H_{\text{bind}}$; a stronger bond has a longer bound lifetime, and thus a smaller $k_{\text{off}}$. This can be manipulated by chemical construction, e.g. opting for hydrogen bonding, coordination bonding, or multiple adjacent binding groups located on the same reversible linker.

On the other hand, the rate of binding $k_{\text{on}}$ depends on, among other things, the chemical potential of the reversible linkers when unbound. One of the key contributions to the chemical potential is the concentration of linkers in the solvent phase. A larger concentration leads to a greater possibility of reversible linker binding events, and thus a larger overall $k_{\text{on}}$. The binding rate $k_{\text{on}}$ also depends on how exquisitely oriented a reversible linker must be, relative to a nearby polymer segment, in order to form a bond. This will depend on the type of chemical interaction they have, as well as steric considerations. The more precisely oriented the reversible linker must be relative to its partner to form a bond, the smaller $k_\mathrm{on}$ will be.

Both $k_{\text{on}}$ and $k_{\text{off}}$ can be adjusted by introducing an activation barrier $\Delta G^{\ddag}$ to reversible linker binding and unbinding. For binding, the total barrier is just $\Delta G^{\ddag}$, while for unbinding it is $\Delta G^{\ddag} - \Delta G_{\text{bind}}$. Thus, the timescale of both $k_{\text{on}}$ and $k_{\text{off}}$ is shifted by the same factor $\propto \exp{(-\Delta G^{\ddag}/RT)}$, yet the ratio of $k_{\text{on}} / k_{\text{off}}$ results in the same $K_{\text{eq}}$ as when the activation barriers are absent. (This is because $K_{\text{eq}} \propto k_{\text{on}} / k_{\text{off}} \propto \exp{(-\Delta G^{\ddag}/RT)} / \exp{[(\Delta G_{\text{bind}} -\Delta G^{\ddag})/RT]}$. The factors of $\exp{(-\Delta G^{\ddag}/RT)}$ cancel to yield $K_{\text{eq}} \propto \exp{(-\Delta G_{\text{bind}}/RT)}$, the intrinsic $K_{\text{eq}}$ of reversible crosslinkers.)

In this way, the rate of reversible crosslink binding/unbinding can be tuned independently from the equilibrium binding strength $K_{\text{eq}}$. An activation barrier can be manipulated by the chemical composition of the reversible crosslinks. For example, bulky side groups could be added to the reversible linker so that there is an initial steric barrier to binding, overcome only when the side group(s) are configured so as to expose the binding site to the polymer backbone.

Thus, a ``weak-binding'' reversible linker, i.e. having a small $K_{\text{eq}}$, can be constructed by \emph{any} choice of ratio $k_{\text{on}} / k_{\text{off}}$ leading to that $K_{\text{eq}}$. Long reversible linker binding lifetimes (i.e. small $k_{\text{off}}$), such as used in Kean et al. 2014, can be offset by comparably small binding rates $k_{\text{on}}$ such that the system is in an overall weak-binding (small $K_{\text{eq}}$) regime. We expect that Kean et al. 2014 are sampling that regime in their system exhibiting strong toughness enhancement yet little change in elasticity. On the other hand, they likely enter into a large-$K_{\text{eq}}$ regime for their strongest reversible linker, and that is why that system exhibits a significantly different modulus relative to the native material.

 As a final point, we remark on how actively exchanging reversible crosslinks can still bear mechanical stress during strain. Reversible crosslinks can continuously propagate and exchange binding partners over the timescale of strain, particularly if the strain is slow relative to the binding/unbinding timescale. Suppose we design the reversible crosslinks to have a small $K_{\text{eq}}$ that is sufficient to yield, \emph{on average}, a few reversible crosslinks bound around each permanent crosslink at any given moment during strain. The macroscopic stress/strain curve represents an average over the ensemble of different local molecular environments, having fluctuations from the mean. For example, over a short interval of microscopic time, some localities in the material might be unbinding and rebinding new reversible crosslinks, while other spots have their reversible links fixed during that time interval. However, at the level of a macroscopic average over all such microscopic environments, the strain experiment will ``feel'' those reversible crosslinks that happen to be bound, and locally reinforcing the material, at any given instant in time. The small $K_{\text{eq}}$ effectively biases these bound linkers so that they are more likely to be clustered around the permanent crosslinks, while a binding/unbinding activation barrier can give the bound linkers enough ``residence time'' to bear some local stress before moving on to a new partner. 

\section{Conclusions}

This work has constructed an equilibrium picture for how reversible crosslinks, allowed to freely diffuse as individual units throughout a polymer network, can toughen the network while maintaining the intrinsic network elasticity. The reversible crosslinks are driven by entropy to recruit to binding positions around the permanent crosslinks in the network, as this leads to the smallest entropy loss for participating polymer chains. The permanent crosslinks are thereby reinforced so that they may bear more local stress before rupturing, leading the network as a whole to be tougher when strained. Since the topology of the polymer network is not significantly altered when the reversible crosslinks recruit around existing permanent links, the material exhibits its original intrinsic elasticity.

From the results and discussion presented here, we can define guidelines for optimal reversible crosslink design. The best reversible crosslink design is one that
\begin{itemize}
	\item{has a low equilibrium constant $K_{\text{eq}}$ of binding to the polymer backbone, so as to selectively cluster to binding positions around permanent crosslinks in the network;}
	\item{is sufficiently strong binding (negative $\Delta H_{\text{bind}}$), or has binding/unbinding activation barriers, so as to productively share local stress near a permanent crosslink;}
	\item{has a binding exchange timescale ($k_{\text{on}}, k_{\text{off}}$) that is shorter than the timescale over which the material will be strained in typical use.}
\end{itemize}
These conditions are easiest to satisfy when the timescale of strain is very long. This regime allows leeway for designing reversible linkers that have a strong binding enthalpy, yet nevertheless have a rearrangement timescale that is still less than the strain timescale. The only condition is to ensure that the binding equilibrium constant of such strong-enthalpy binders is sufficiently low so that they entropically recruit around the permanent crosslinks. Adjusting the reversible crosslink concentration is a simple first tactic to reaching this weak-binding low-$K_{\text{eq}}$ regime, while the chemistry of the reversible linkers can be used to tune the binding enthalpy $\Delta H_{\text{bind}}$ and on/off rates $k_{\text{on}}$ and $k_{\text{off}}$.

A more challenging scenario is a material where strain is occurring on ``short'' (mesoscopic or microscopic) timescales. Here, the window for making reversible linkers that can bear local stress yet rearrange below the strain timescale becomes narrow. The rearrangement kinetics of the reversible crosslinks become dominant in the behaviour of the material during strain, which we leave to future study.

A key question arising out of this study is how microscopically enhancing the strength of crosslinks in a polymer network leads the material to be macroscopically tougher by an even larger factor. Our SCFT network model can be used to examine the length scale over which polymer strands break as a network is strained to rupture point, which may play a role in this. This length scale additionally has predictive power for assessing the macroscopic behaviour of the material at the tip of a crack during rupture. A systematic examination of toughness enhancement in different network topologies will also be undertaken. 

\section{Acknowledgments}

This research has been performed within the framework of the 4TU.High-Tech Materials research programme `New Horizons in designer materials' (www.4tu.nl/htm). We are grateful for compute time obtained from the Netherlands Organisation for Scientific Research (NWO) SURFsara Pilot Program (Grant No. 15582). Additional thanks to Rint Sijbesma, Rob Jack, Yuval Mulla, and Gijsje Koenderink for valuable discussions. 
\\
\\
\noindent \textbf{Supporting Information.} Mathematical derivations of key equations in main text, molecular dynamics simulation details, and self-consistent field calculation details.

\bibliography{main}

\begin{thebibliography}{60}%
\makeatletter
\providecommand \@ifxundefined [1]{%
 \@ifx{#1\undefined}
}%
\providecommand \@ifnum [1]{%
 \ifnum #1\expandafter \@firstoftwo
 \else \expandafter \@secondoftwo
 \fi
}%
\providecommand \@ifx [1]{%
 \ifx #1\expandafter \@firstoftwo
 \else \expandafter \@secondoftwo
 \fi
}%
\providecommand \natexlab [1]{#1}%
\providecommand \enquote  [1]{``#1''}%
\providecommand \bibnamefont  [1]{#1}%
\providecommand \bibfnamefont [1]{#1}%
\providecommand \citenamefont [1]{#1}%
\providecommand \href@noop [0]{\@secondoftwo}%
\providecommand \href [0]{\begingroup \@sanitize@url \@href}%
\providecommand \@href[1]{\@@startlink{#1}\@@href}%
\providecommand \@@href[1]{\endgroup#1\@@endlink}%
\providecommand \@sanitize@url [0]{\catcode `\\12\catcode `\$12\catcode
  `\&12\catcode `\#12\catcode `\^12\catcode `\_12\catcode `\%12\relax}%
\providecommand \@@startlink[1]{}%
\providecommand \@@endlink[0]{}%
\providecommand \url  [0]{\begingroup\@sanitize@url \@url }%
\providecommand \@url [1]{\endgroup\@href {#1}{\urlprefix }}%
\providecommand \urlprefix  [0]{URL }%
\providecommand \Eprint [0]{\href }%
\providecommand \doibase [0]{http://dx.doi.org/}%
\providecommand \selectlanguage [0]{\@gobble}%
\providecommand \bibinfo  [0]{\@secondoftwo}%
\providecommand \bibfield  [0]{\@secondoftwo}%
\providecommand \translation [1]{[#1]}%
\providecommand \BibitemOpen [0]{}%
\providecommand \bibitemStop [0]{}%
\providecommand \bibitemNoStop [0]{.\EOS\space}%
\providecommand \EOS [0]{\spacefactor3000\relax}%
\providecommand \BibitemShut  [1]{\csname bibitem#1\endcsname}%
\let\auto@bib@innerbib\@empty
\bibitem [{\citenamefont {Lake}\ and\ \citenamefont
  {Thomas}(1967)}]{Lake:1967ec}%
  \BibitemOpen
  \bibfield  {author} {\bibinfo {author} {\bibfnamefont {G.~J.}\ \bibnamefont
  {Lake}}\ and\ \bibinfo {author} {\bibfnamefont {A.~G.}\ \bibnamefont
  {Thomas}},\ }\href@noop {} {\bibfield  {journal} {\bibinfo  {journal}
  {Proceedings of the Royal Society of London Series A-Mathematical and
  Physical Sciences}\ }\textbf {\bibinfo {volume} {300}},\ \bibinfo {pages}
  {108} (\bibinfo {year} {1967})}\BibitemShut {NoStop}%
\bibitem [{\citenamefont {Ligoure}\ and\ \citenamefont
  {Mora}(2013)}]{Ligoure:2013bd}%
  \BibitemOpen
  \bibfield  {author} {\bibinfo {author} {\bibfnamefont {C.}~\bibnamefont
  {Ligoure}}\ and\ \bibinfo {author} {\bibfnamefont {S.}~\bibnamefont {Mora}},\
  }\href@noop {} {\bibfield  {journal} {\bibinfo  {journal} {Rheologica Acta}\
  }\textbf {\bibinfo {volume} {52}},\ \bibinfo {pages} {91} (\bibinfo {year}
  {2013})}\BibitemShut {NoStop}%
\bibitem [{\citenamefont {Creton}(2017)}]{Creton:2017bd}%
  \BibitemOpen
  \bibfield  {author} {\bibinfo {author} {\bibfnamefont {C.}~\bibnamefont
  {Creton}},\ }\href@noop {} {\bibfield  {journal} {\bibinfo  {journal}
  {Macromolecules}\ }\textbf {\bibinfo {volume} {50}},\ \bibinfo {pages} {8297}
  (\bibinfo {year} {2017})}\BibitemShut {NoStop}%
\bibitem [{\citenamefont {Creton}\ and\ \citenamefont
  {Ciccotti}(2016)}]{Creton:2016fr}%
  \BibitemOpen
  \bibfield  {author} {\bibinfo {author} {\bibfnamefont {C.}~\bibnamefont
  {Creton}}\ and\ \bibinfo {author} {\bibfnamefont {M.}~\bibnamefont
  {Ciccotti}},\ }\href@noop {} {\bibfield  {journal} {\bibinfo  {journal}
  {Reports on Progress in Physics}\ }\textbf {\bibinfo {volume} {79}},\
  \bibinfo {pages} {046601} (\bibinfo {year} {2016})}\BibitemShut {NoStop}%
\bibitem [{\citenamefont {Linga}, \citenamefont {Ballone},\ and\ \citenamefont
  {Hansen}(2015)}]{Linga:2015cy}%
  \BibitemOpen
  \bibfield  {author} {\bibinfo {author} {\bibfnamefont {G.}~\bibnamefont
  {Linga}}, \bibinfo {author} {\bibfnamefont {P.}~\bibnamefont {Ballone}}, \
  and\ \bibinfo {author} {\bibfnamefont {A.}~\bibnamefont {Hansen}},\
  }\href@noop {} {\bibfield  {journal} {\bibinfo  {journal} {Physical Review
  E}\ }\textbf {\bibinfo {volume} {92}},\ \bibinfo {pages} {1065} (\bibinfo
  {year} {2015})}\BibitemShut {NoStop}%
\bibitem [{\citenamefont {Kim}\ and\ \citenamefont {Wool}(1983)}]{Kim:1983em}%
  \BibitemOpen
  \bibfield  {author} {\bibinfo {author} {\bibfnamefont {Y.~H.}\ \bibnamefont
  {Kim}}\ and\ \bibinfo {author} {\bibfnamefont {R.~P.}\ \bibnamefont {Wool}},\
  }\href@noop {} {\bibfield  {journal} {\bibinfo  {journal} {Macromolecules}\
  }\textbf {\bibinfo {volume} {16}},\ \bibinfo {pages} {1115} (\bibinfo {year}
  {1983})}\BibitemShut {NoStop}%
\bibitem [{\citenamefont {Zwaag}\ and\ \citenamefont
  {Brinkman}(2015)}]{Zwaag:2007ue}%
  \BibitemOpen
  \bibfield  {author} {\bibinfo {author} {\bibfnamefont {S.~v.~d.}\
  \bibnamefont {Zwaag}}\ and\ \bibinfo {author} {\bibfnamefont
  {E.}~\bibnamefont {Brinkman}},\ }\href@noop {} {\emph {\bibinfo {title}
  {{Self healing materials}}}}\ (\bibinfo  {publisher} {IOS Press},\ \bibinfo
  {address} {Amsterdam, The Netherlands},\ \bibinfo {year} {2015})\BibitemShut
  {NoStop}%
\bibitem [{\citenamefont {Wool}(2008)}]{Wool:2008gc}%
  \BibitemOpen
  \bibfield  {author} {\bibinfo {author} {\bibfnamefont {R.~P.}\ \bibnamefont
  {Wool}},\ }\href@noop {} {\bibfield  {journal} {\bibinfo  {journal} {Soft
  Matter}\ }\textbf {\bibinfo {volume} {4}},\ \bibinfo {pages} {400} (\bibinfo
  {year} {2008})}\BibitemShut {NoStop}%
\bibitem [{\citenamefont {Stukalin}\ \emph {et~al.}(2013)\citenamefont
  {Stukalin}, \citenamefont {Cai}, \citenamefont {Kumar}, \citenamefont
  {Leibler},\ and\ \citenamefont {Rubinstein}}]{Stukalin:2013ju}%
  \BibitemOpen
  \bibfield  {author} {\bibinfo {author} {\bibfnamefont {E.~B.}\ \bibnamefont
  {Stukalin}}, \bibinfo {author} {\bibfnamefont {L.-H.}\ \bibnamefont {Cai}},
  \bibinfo {author} {\bibfnamefont {N.~A.}\ \bibnamefont {Kumar}}, \bibinfo
  {author} {\bibfnamefont {L.}~\bibnamefont {Leibler}}, \ and\ \bibinfo
  {author} {\bibfnamefont {M.}~\bibnamefont {Rubinstein}},\ }\href@noop {}
  {\bibfield  {journal} {\bibinfo  {journal} {Macromolecules}\ }\textbf
  {\bibinfo {volume} {46}},\ \bibinfo {pages} {7525} (\bibinfo {year}
  {2013})}\BibitemShut {NoStop}%
\bibitem [{\citenamefont {Amaral}\ and\ \citenamefont
  {Pasparakis}(2017)}]{Amaral:2017bi}%
  \BibitemOpen
  \bibfield  {author} {\bibinfo {author} {\bibfnamefont {A.~J.~R.}\
  \bibnamefont {Amaral}}\ and\ \bibinfo {author} {\bibfnamefont
  {G.}~\bibnamefont {Pasparakis}},\ }\href@noop {} {\bibfield  {journal}
  {\bibinfo  {journal} {Polymer Chemistry}\ }\textbf {\bibinfo {volume} {8}},\
  \bibinfo {pages} {6464} (\bibinfo {year} {2017})}\BibitemShut {NoStop}%
\bibitem [{\citenamefont {Montarnal}\ \emph {et~al.}(2011)\citenamefont
  {Montarnal}, \citenamefont {Capelot}, \citenamefont {Tournilhac},\ and\
  \citenamefont {Leibler}}]{Montarnal:2011cr}%
  \BibitemOpen
  \bibfield  {author} {\bibinfo {author} {\bibfnamefont {D.}~\bibnamefont
  {Montarnal}}, \bibinfo {author} {\bibfnamefont {M.}~\bibnamefont {Capelot}},
  \bibinfo {author} {\bibfnamefont {F.}~\bibnamefont {Tournilhac}}, \ and\
  \bibinfo {author} {\bibfnamefont {L.}~\bibnamefont {Leibler}},\ }\href@noop
  {} {\bibfield  {journal} {\bibinfo  {journal} {Science}\ }\textbf {\bibinfo
  {volume} {334}},\ \bibinfo {pages} {965} (\bibinfo {year}
  {2011})}\BibitemShut {NoStop}%
\bibitem [{\citenamefont {Kloxin}\ and\ \citenamefont
  {Bowman}(2013)}]{Kloxin:2013hu}%
  \BibitemOpen
  \bibfield  {author} {\bibinfo {author} {\bibfnamefont {C.~J.}\ \bibnamefont
  {Kloxin}}\ and\ \bibinfo {author} {\bibfnamefont {C.~N.}\ \bibnamefont
  {Bowman}},\ }\href@noop {} {\bibfield  {journal} {\bibinfo  {journal}
  {Chemical Society Reviews}\ }\textbf {\bibinfo {volume} {42}},\ \bibinfo
  {pages} {7161} (\bibinfo {year} {2013})}\BibitemShut {NoStop}%
\bibitem [{\citenamefont {Narita}\ \emph {et~al.}(2013)\citenamefont {Narita},
  \citenamefont {Mayumi}, \citenamefont {Ducouret},\ and\ \citenamefont
  {H{\'e}braud}}]{Narita:2013kt}%
  \BibitemOpen
  \bibfield  {author} {\bibinfo {author} {\bibfnamefont {T.}~\bibnamefont
  {Narita}}, \bibinfo {author} {\bibfnamefont {K.}~\bibnamefont {Mayumi}},
  \bibinfo {author} {\bibfnamefont {G.}~\bibnamefont {Ducouret}}, \ and\
  \bibinfo {author} {\bibfnamefont {P.}~\bibnamefont {H{\'e}braud}},\
  }\href@noop {} {\bibfield  {journal} {\bibinfo  {journal} {Macromolecules}\
  }\textbf {\bibinfo {volume} {46}},\ \bibinfo {pages} {4174} (\bibinfo {year}
  {2013})}\BibitemShut {NoStop}%
\bibitem [{\citenamefont {Kean}\ \emph {et~al.}(2014)\citenamefont {Kean},
  \citenamefont {Hawk}, \citenamefont {Lin}, \citenamefont {Zhao},
  \citenamefont {Sijbesma},\ and\ \citenamefont {Craig}}]{Kean:2014hh}%
  \BibitemOpen
  \bibfield  {author} {\bibinfo {author} {\bibfnamefont {Z.~S.}\ \bibnamefont
  {Kean}}, \bibinfo {author} {\bibfnamefont {J.~L.}\ \bibnamefont {Hawk}},
  \bibinfo {author} {\bibfnamefont {S.}~\bibnamefont {Lin}}, \bibinfo {author}
  {\bibfnamefont {X.}~\bibnamefont {Zhao}}, \bibinfo {author} {\bibfnamefont
  {R.~P.}\ \bibnamefont {Sijbesma}}, \ and\ \bibinfo {author} {\bibfnamefont
  {S.~L.}\ \bibnamefont {Craig}},\ }\href@noop {} {\bibfield  {journal}
  {\bibinfo  {journal} {Advanced Materials}\ }\textbf {\bibinfo {volume}
  {26}},\ \bibinfo {pages} {6013} (\bibinfo {year} {2014})}\BibitemShut
  {NoStop}%
\bibitem [{\citenamefont {Li}\ \emph {et~al.}(2015)\citenamefont {Li},
  \citenamefont {Barrett}, \citenamefont {Messersmith},\ and\ \citenamefont
  {Holten-Andersen}}]{Li:2015bm}%
  \BibitemOpen
  \bibfield  {author} {\bibinfo {author} {\bibfnamefont {Q.}~\bibnamefont
  {Li}}, \bibinfo {author} {\bibfnamefont {D.~G.}\ \bibnamefont {Barrett}},
  \bibinfo {author} {\bibfnamefont {P.~B.}\ \bibnamefont {Messersmith}}, \ and\
  \bibinfo {author} {\bibfnamefont {N.}~\bibnamefont {Holten-Andersen}},\
  }\href@noop {} {\bibfield  {journal} {\bibinfo  {journal} {ACS Nano}\
  }\textbf {\bibinfo {volume} {10}},\ \bibinfo {pages} {1317} (\bibinfo {year}
  {2015})}\BibitemShut {NoStop}%
\bibitem [{\citenamefont {Neal}, \citenamefont {Mozhdehi},\ and\ \citenamefont
  {Guan}(2015)}]{Neal:2015ju}%
  \BibitemOpen
  \bibfield  {author} {\bibinfo {author} {\bibfnamefont {J.~A.}\ \bibnamefont
  {Neal}}, \bibinfo {author} {\bibfnamefont {D.}~\bibnamefont {Mozhdehi}}, \
  and\ \bibinfo {author} {\bibfnamefont {Z.}~\bibnamefont {Guan}},\ }\href@noop
  {} {\bibfield  {journal} {\bibinfo  {journal} {Journal of the American
  Chemical Society}\ }\textbf {\bibinfo {volume} {137}},\ \bibinfo {pages}
  {4846} (\bibinfo {year} {2015})}\BibitemShut {NoStop}%
\bibitem [{\citenamefont {Kurniawan}\ \emph {et~al.}(2016)\citenamefont
  {Kurniawan}, \citenamefont {Vos}, \citenamefont {Biebricher}, \citenamefont
  {Wuite}, \citenamefont {Peterman},\ and\ \citenamefont
  {Koenderink}}]{Kurniawan:2016hp}%
  \BibitemOpen
  \bibfield  {author} {\bibinfo {author} {\bibfnamefont {N.~A.}\ \bibnamefont
  {Kurniawan}}, \bibinfo {author} {\bibfnamefont {B.~E.}\ \bibnamefont {Vos}},
  \bibinfo {author} {\bibfnamefont {A.}~\bibnamefont {Biebricher}}, \bibinfo
  {author} {\bibfnamefont {G.~J.~L.}\ \bibnamefont {Wuite}}, \bibinfo {author}
  {\bibfnamefont {E.~J.~G.}\ \bibnamefont {Peterman}}, \ and\ \bibinfo {author}
  {\bibfnamefont {G.~H.}\ \bibnamefont {Koenderink}},\ }\href@noop {}
  {\bibfield  {journal} {\bibinfo  {journal} {Biophysical Journal}\ }\textbf
  {\bibinfo {volume} {111}},\ \bibinfo {pages} {1026} (\bibinfo {year}
  {2016})}\BibitemShut {NoStop}%
\bibitem [{\citenamefont {Imbernon}, \citenamefont {Norvez},\ and\
  \citenamefont {Leibler}(2016)}]{Imbernon:2016kb}%
  \BibitemOpen
  \bibfield  {author} {\bibinfo {author} {\bibfnamefont {L.}~\bibnamefont
  {Imbernon}}, \bibinfo {author} {\bibfnamefont {S.}~\bibnamefont {Norvez}}, \
  and\ \bibinfo {author} {\bibfnamefont {L.}~\bibnamefont {Leibler}},\
  }\href@noop {} {\bibfield  {journal} {\bibinfo  {journal} {Macromolecules}\
  }\textbf {\bibinfo {volume} {49}},\ \bibinfo {pages} {2172} (\bibinfo {year}
  {2016})}\BibitemShut {NoStop}%
\bibitem [{\citenamefont {White}\ and\ \citenamefont
  {Broer}(2015)}]{White:2015if}%
  \BibitemOpen
  \bibfield  {author} {\bibinfo {author} {\bibfnamefont {T.~J.}\ \bibnamefont
  {White}}\ and\ \bibinfo {author} {\bibfnamefont {D.~J.}\ \bibnamefont
  {Broer}},\ }\href@noop {} {\bibfield  {journal} {\bibinfo  {journal} {Nature
  Publishing Group}\ }\textbf {\bibinfo {volume} {14}},\ \bibinfo {pages}
  {1087} (\bibinfo {year} {2015})}\BibitemShut {NoStop}%
\bibitem [{\citenamefont {Hoy}\ and\ \citenamefont
  {Fredrickson}(2009)}]{Hoy:2009gm}%
  \BibitemOpen
  \bibfield  {author} {\bibinfo {author} {\bibfnamefont {R.~S.}\ \bibnamefont
  {Hoy}}\ and\ \bibinfo {author} {\bibfnamefont {G.~H.}\ \bibnamefont
  {Fredrickson}},\ }\href@noop {} {\bibfield  {journal} {\bibinfo  {journal}
  {Journal of Chemical Physics}\ }\textbf {\bibinfo {volume} {131}},\ \bibinfo
  {pages} {224902} (\bibinfo {year} {2009})}\BibitemShut {NoStop}%
\bibitem [{\citenamefont {Sprakel}\ \emph {et~al.}(2009)\citenamefont
  {Sprakel}, \citenamefont {Spruijt}, \citenamefont {van~der Gucht},
  \citenamefont {Padding},\ and\ \citenamefont {Briels}}]{Sprakel:2009gs}%
  \BibitemOpen
  \bibfield  {author} {\bibinfo {author} {\bibfnamefont {J.}~\bibnamefont
  {Sprakel}}, \bibinfo {author} {\bibfnamefont {E.}~\bibnamefont {Spruijt}},
  \bibinfo {author} {\bibfnamefont {J.}~\bibnamefont {van~der Gucht}}, \bibinfo
  {author} {\bibfnamefont {J.~T.}\ \bibnamefont {Padding}}, \ and\ \bibinfo
  {author} {\bibfnamefont {W.~J.}\ \bibnamefont {Briels}},\ }\href@noop {}
  {\bibfield  {journal} {\bibinfo  {journal} {Soft Matter}\ }\textbf {\bibinfo
  {volume} {5}},\ \bibinfo {pages} {4748} (\bibinfo {year} {2009})}\BibitemShut
  {NoStop}%
\bibitem [{\citenamefont {Indei}\ and\ \citenamefont
  {Takimoto}(2010)}]{Indei:2010kl}%
  \BibitemOpen
  \bibfield  {author} {\bibinfo {author} {\bibfnamefont {T.}~\bibnamefont
  {Indei}}\ and\ \bibinfo {author} {\bibfnamefont {J.-i.}\ \bibnamefont
  {Takimoto}},\ }\href@noop {} {\bibfield  {journal} {\bibinfo  {journal} {The
  Journal of Chemical Physics}\ }\textbf {\bibinfo {volume} {133}},\ \bibinfo
  {pages} {194902} (\bibinfo {year} {2010})}\BibitemShut {NoStop}%
\bibitem [{\citenamefont {Nabavi}, \citenamefont {Fratzl},\ and\ \citenamefont
  {Hartmann}(2015)}]{Nabavi:2015ki}%
  \BibitemOpen
  \bibfield  {author} {\bibinfo {author} {\bibfnamefont {S.~S.}\ \bibnamefont
  {Nabavi}}, \bibinfo {author} {\bibfnamefont {P.}~\bibnamefont {Fratzl}}, \
  and\ \bibinfo {author} {\bibfnamefont {M.~A.}\ \bibnamefont {Hartmann}},\
  }\href@noop {} {\bibfield  {journal} {\bibinfo  {journal} {Physical Review
  E}\ }\textbf {\bibinfo {volume} {91}},\ \bibinfo {pages} {032603} (\bibinfo
  {year} {2015})}\BibitemShut {NoStop}%
\bibitem [{\citenamefont {West}\ and\ \citenamefont
  {Kindt}(2015)}]{West:2015fe}%
  \BibitemOpen
  \bibfield  {author} {\bibinfo {author} {\bibfnamefont {A.}~\bibnamefont
  {West}}\ and\ \bibinfo {author} {\bibfnamefont {J.~T.}\ \bibnamefont
  {Kindt}},\ }\href@noop {} {\bibfield  {journal} {\bibinfo  {journal}
  {Macromolecular Theory and Simulations}\ }\textbf {\bibinfo {volume} {24}},\
  \bibinfo {pages} {208} (\bibinfo {year} {2015})}\BibitemShut {NoStop}%
\bibitem [{\citenamefont {Gordievskaya}, \citenamefont {Rumyantsev},\ and\
  \citenamefont {Kramarenko}(2016)}]{Gordievskaya:2016es}%
  \BibitemOpen
  \bibfield  {author} {\bibinfo {author} {\bibfnamefont {Y.~D.}\ \bibnamefont
  {Gordievskaya}}, \bibinfo {author} {\bibfnamefont {A.~M.}\ \bibnamefont
  {Rumyantsev}}, \ and\ \bibinfo {author} {\bibfnamefont {E.~Y.}\ \bibnamefont
  {Kramarenko}},\ }\href@noop {} {\bibfield  {journal} {\bibinfo  {journal}
  {Journal of Chemical Physics}\ }\textbf {\bibinfo {volume} {144}},\ \bibinfo
  {pages} {184902} (\bibinfo {year} {2016})}\BibitemShut {NoStop}%
\bibitem [{\citenamefont {Amin}, \citenamefont {Likhtman},\ and\ \citenamefont
  {Wang}(2016)}]{Amin:2016je}%
  \BibitemOpen
  \bibfield  {author} {\bibinfo {author} {\bibfnamefont {D.}~\bibnamefont
  {Amin}}, \bibinfo {author} {\bibfnamefont {A.~E.}\ \bibnamefont {Likhtman}},
  \ and\ \bibinfo {author} {\bibfnamefont {Z.}~\bibnamefont {Wang}},\
  }\href@noop {} {\bibfield  {journal} {\bibinfo  {journal} {Macromolecules}\
  }\textbf {\bibinfo {volume} {49}},\ \bibinfo {pages} {7510} (\bibinfo {year}
  {2016})}\BibitemShut {NoStop}%
\bibitem [{\citenamefont {Meng}, \citenamefont {Pritchard},\ and\ \citenamefont
  {Terentjev}(2016)}]{Meng:2016hs}%
  \BibitemOpen
  \bibfield  {author} {\bibinfo {author} {\bibfnamefont {F.}~\bibnamefont
  {Meng}}, \bibinfo {author} {\bibfnamefont {R.~H.}\ \bibnamefont {Pritchard}},
  \ and\ \bibinfo {author} {\bibfnamefont {E.~M.}\ \bibnamefont {Terentjev}},\
  }\href@noop {} {\bibfield  {journal} {\bibinfo  {journal} {Macromolecules}\
  }\textbf {\bibinfo {volume} {49}},\ \bibinfo {pages} {2843} (\bibinfo {year}
  {2016})}\BibitemShut {NoStop}%
\bibitem [{\citenamefont {Yu}\ \emph {et~al.}(2016)\citenamefont {Yu},
  \citenamefont {Shi}, \citenamefont {Li}, \citenamefont {Jabour},
  \citenamefont {Yang}, \citenamefont {Dunn}, \citenamefont {Wang},\ and\
  \citenamefont {Qi}}]{Yu:2016dl}%
  \BibitemOpen
  \bibfield  {author} {\bibinfo {author} {\bibfnamefont {K.}~\bibnamefont
  {Yu}}, \bibinfo {author} {\bibfnamefont {Q.}~\bibnamefont {Shi}}, \bibinfo
  {author} {\bibfnamefont {H.}~\bibnamefont {Li}}, \bibinfo {author}
  {\bibfnamefont {J.}~\bibnamefont {Jabour}}, \bibinfo {author} {\bibfnamefont
  {H.}~\bibnamefont {Yang}}, \bibinfo {author} {\bibfnamefont {M.~L.}\
  \bibnamefont {Dunn}}, \bibinfo {author} {\bibfnamefont {T.}~\bibnamefont
  {Wang}}, \ and\ \bibinfo {author} {\bibfnamefont {H.~J.}\ \bibnamefont
  {Qi}},\ }\href@noop {} {\bibfield  {journal} {\bibinfo  {journal} {Journal of
  the Mechanics and Physics of Solids}\ }\textbf {\bibinfo {volume} {94}},\
  \bibinfo {pages} {1} (\bibinfo {year} {2016})}\BibitemShut {NoStop}%
\bibitem [{\citenamefont {Brighenti}\ and\ \citenamefont
  {Vernerey}(2017)}]{Brighenti:2017ff}%
  \BibitemOpen
  \bibfield  {author} {\bibinfo {author} {\bibfnamefont {R.}~\bibnamefont
  {Brighenti}}\ and\ \bibinfo {author} {\bibfnamefont {F.~J.}\ \bibnamefont
  {Vernerey}},\ }\href@noop {} {\bibfield  {journal} {\bibinfo  {journal}
  {Composites Part B: Engineering}\ }\textbf {\bibinfo {volume} {115}},\
  \bibinfo {pages} {257} (\bibinfo {year} {2017})}\BibitemShut {NoStop}%
\bibitem [{\citenamefont {Vernerey}, \citenamefont {Long},\ and\ \citenamefont
  {Brighenti}(2017)}]{Vernerey:2017gt}%
  \BibitemOpen
  \bibfield  {author} {\bibinfo {author} {\bibfnamefont {F.~J.}\ \bibnamefont
  {Vernerey}}, \bibinfo {author} {\bibfnamefont {R.}~\bibnamefont {Long}}, \
  and\ \bibinfo {author} {\bibfnamefont {R.}~\bibnamefont {Brighenti}},\
  }\href@noop {} {\bibfield  {journal} {\bibinfo  {journal} {Journal of the
  Mechanics and Physics of Solids}\ }\textbf {\bibinfo {volume} {107}},\
  \bibinfo {pages} {1} (\bibinfo {year} {2017})}\BibitemShut {NoStop}%
\bibitem [{\citenamefont {Oyarz{\'u}n}\ and\ \citenamefont
  {Mognetti}(2018{\natexlab{a}})}]{Oyarzun:2018hz}%
  \BibitemOpen
  \bibfield  {author} {\bibinfo {author} {\bibfnamefont {B.}~\bibnamefont
  {Oyarz{\'u}n}}\ and\ \bibinfo {author} {\bibfnamefont {B.~M.}\ \bibnamefont
  {Mognetti}},\ }\href@noop {} {\bibfield  {journal} {\bibinfo  {journal}
  {Journal of Chemical Physics}\ }\textbf {\bibinfo {volume} {148}},\ \bibinfo
  {pages} {114110} (\bibinfo {year} {2018}{\natexlab{a}})}\BibitemShut
  {NoStop}%
\bibitem [{\citenamefont {Oyarz{\'u}n}\ and\ \citenamefont
  {Mognetti}(2018{\natexlab{b}})}]{Oyarzun:2018tk}%
  \BibitemOpen
  \bibfield  {author} {\bibinfo {author} {\bibfnamefont {B.}~\bibnamefont
  {Oyarz{\'u}n}}\ and\ \bibinfo {author} {\bibfnamefont {B.~M.}\ \bibnamefont
  {Mognetti}},\ }\href@noop {} {\bibfield  {journal} {\bibinfo  {journal} {Mol.
  Phys. (in press)}\ ,\ \bibinfo {pages} {1}} (\bibinfo {year}
  {2018}{\natexlab{b}})}\BibitemShut {NoStop}%
\bibitem [{\citenamefont {Parada}\ and\ \citenamefont
  {Zhao}(2018)}]{Parada:2018dk}%
  \BibitemOpen
  \bibfield  {author} {\bibinfo {author} {\bibfnamefont {G.~A.}\ \bibnamefont
  {Parada}}\ and\ \bibinfo {author} {\bibfnamefont {X.}~\bibnamefont {Zhao}},\
  }\href@noop {} {\bibfield  {journal} {\bibinfo  {journal} {Soft Matter}\
  }\textbf {\bibinfo {volume} {14}},\ \bibinfo {pages} {5186} (\bibinfo {year}
  {2018})}\BibitemShut {NoStop}%
\bibitem [{\citenamefont {Mayumi}\ \emph {et~al.}(2016)\citenamefont {Mayumi},
  \citenamefont {Guo}, \citenamefont {Narita}, \citenamefont {Hui},\ and\
  \citenamefont {Creton}}]{Mayumi:2016kr}%
  \BibitemOpen
  \bibfield  {author} {\bibinfo {author} {\bibfnamefont {K.}~\bibnamefont
  {Mayumi}}, \bibinfo {author} {\bibfnamefont {J.}~\bibnamefont {Guo}},
  \bibinfo {author} {\bibfnamefont {T.}~\bibnamefont {Narita}}, \bibinfo
  {author} {\bibfnamefont {C.~Y.}\ \bibnamefont {Hui}}, \ and\ \bibinfo
  {author} {\bibfnamefont {C.}~\bibnamefont {Creton}},\ }\href@noop {}
  {\bibfield  {journal} {\bibinfo  {journal} {Extreme Mechanics Letters}\
  }\textbf {\bibinfo {volume} {6}},\ \bibinfo {pages} {52} (\bibinfo {year}
  {2016})}\BibitemShut {NoStop}%
\bibitem [{\citenamefont {Albrecht}\ and\ \citenamefont {van
  Koten}(2001)}]{Albrecht:2001um}%
  \BibitemOpen
  \bibfield  {author} {\bibinfo {author} {\bibfnamefont {M.}~\bibnamefont
  {Albrecht}}\ and\ \bibinfo {author} {\bibfnamefont {G.}~\bibnamefont {van
  Koten}},\ }\href@noop {} {\bibfield  {journal} {\bibinfo  {journal}
  {Angewandte Chemie International Edition}\ }\textbf {\bibinfo {volume}
  {40}},\ \bibinfo {pages} {3750} (\bibinfo {year} {2001})}\BibitemShut
  {NoStop}%
\bibitem [{\citenamefont {Baxandall}(1989)}]{Baxandall:1989kw}%
  \BibitemOpen
  \bibfield  {author} {\bibinfo {author} {\bibfnamefont {L.~G.}\ \bibnamefont
  {Baxandall}},\ }\href@noop {} {\bibfield  {journal} {\bibinfo  {journal}
  {Macromolecules}\ }\textbf {\bibinfo {volume} {22}},\ \bibinfo {pages} {1982}
  (\bibinfo {year} {1989})}\BibitemShut {NoStop}%
\bibitem [{\citenamefont {Leibler}, \citenamefont {Rubinstein},\ and\
  \citenamefont {Colby}(1991)}]{Leibler:1991hj}%
  \BibitemOpen
  \bibfield  {author} {\bibinfo {author} {\bibfnamefont {L.}~\bibnamefont
  {Leibler}}, \bibinfo {author} {\bibfnamefont {M.}~\bibnamefont {Rubinstein}},
  \ and\ \bibinfo {author} {\bibfnamefont {R.~H.}\ \bibnamefont {Colby}},\
  }\href@noop {} {\bibfield  {journal} {\bibinfo  {journal} {Macromolecules}\
  }\textbf {\bibinfo {volume} {24}},\ \bibinfo {pages} {4701} (\bibinfo {year}
  {1991})}\BibitemShut {NoStop}%
\bibitem [{\citenamefont {Hackelbusch}\ \emph {et~al.}(2013)\citenamefont
  {Hackelbusch}, \citenamefont {Rossow}, \citenamefont {van Assenbergh},\ and\
  \citenamefont {Seiffert}}]{Hackelbusch:2013kq}%
  \BibitemOpen
  \bibfield  {author} {\bibinfo {author} {\bibfnamefont {S.}~\bibnamefont
  {Hackelbusch}}, \bibinfo {author} {\bibfnamefont {T.}~\bibnamefont {Rossow}},
  \bibinfo {author} {\bibfnamefont {P.}~\bibnamefont {van Assenbergh}}, \ and\
  \bibinfo {author} {\bibfnamefont {S.}~\bibnamefont {Seiffert}},\ }\href@noop
  {} {\bibfield  {journal} {\bibinfo  {journal} {Macromolecules}\ }\textbf
  {\bibinfo {volume} {46}},\ \bibinfo {pages} {6273} (\bibinfo {year}
  {2013})}\BibitemShut {NoStop}%
\bibitem [{\citenamefont {Mulla}\ and\ \citenamefont
  {Koenderink}(2018)}]{Mulla:2018uh}%
  \BibitemOpen
  \bibfield  {author} {\bibinfo {author} {\bibfnamefont {Y.}~\bibnamefont
  {Mulla}}\ and\ \bibinfo {author} {\bibfnamefont {G.~H.}\ \bibnamefont
  {Koenderink}},\ }\href@noop {} {\bibfield  {journal} {\bibinfo  {journal}
  {arXiv:1805.12431v1 [cond-mat.soft]}\ ,\ \bibinfo {pages} {1}} (\bibinfo
  {year} {2018})}\BibitemShut {NoStop}%
\bibitem [{\citenamefont {Mulla}\ \emph {et~al.}(2018)\citenamefont {Mulla},
  \citenamefont {Oliveri}, \citenamefont {Overvelde},\ and\ \citenamefont
  {Koenderink}}]{Mulla:2018ga}%
  \BibitemOpen
  \bibfield  {author} {\bibinfo {author} {\bibfnamefont {Y.}~\bibnamefont
  {Mulla}}, \bibinfo {author} {\bibfnamefont {G.}~\bibnamefont {Oliveri}},
  \bibinfo {author} {\bibfnamefont {J.~T.~B.}\ \bibnamefont {Overvelde}}, \
  and\ \bibinfo {author} {\bibfnamefont {G.~H.}\ \bibnamefont {Koenderink}},\
  }\href@noop {} {\bibfield  {journal} {\bibinfo  {journal} {Physical Review
  Letters}\ }\textbf {\bibinfo {volume} {120}},\ \bibinfo {pages} {268002}
  (\bibinfo {year} {2018})}\BibitemShut {NoStop}%
\bibitem [{\citenamefont {Foyart}\ \emph {et~al.}(2016)\citenamefont {Foyart},
  \citenamefont {Ligoure}, \citenamefont {Mora},\ and\ \citenamefont
  {Ramos}}]{Foyart:2016hu}%
  \BibitemOpen
  \bibfield  {author} {\bibinfo {author} {\bibfnamefont {G.}~\bibnamefont
  {Foyart}}, \bibinfo {author} {\bibfnamefont {C.}~\bibnamefont {Ligoure}},
  \bibinfo {author} {\bibfnamefont {S.}~\bibnamefont {Mora}}, \ and\ \bibinfo
  {author} {\bibfnamefont {L.}~\bibnamefont {Ramos}},\ }\href@noop {}
  {\bibfield  {journal} {\bibinfo  {journal} {ACS Macro Letters}\ }\textbf
  {\bibinfo {volume} {5}},\ \bibinfo {pages} {1080} (\bibinfo {year}
  {2016})}\BibitemShut {NoStop}%
\bibitem [{\citenamefont {Treloar}(1942)}]{Treloar:1942}%
  \BibitemOpen
  \bibfield  {author} {\bibinfo {author} {\bibfnamefont {L.~R.~G.}\
  \bibnamefont {Treloar}},\ }\href@noop {} {\bibfield  {journal} {\bibinfo
  {journal} {Reports on Progress in Physics}\ }\textbf {\bibinfo {volume}
  {9}},\ \bibinfo {pages} {113} (\bibinfo {year} {1942})}\BibitemShut {NoStop}%
\bibitem [{\citenamefont {James}(1947)}]{JAMES:1947hp}%
  \BibitemOpen
  \bibfield  {author} {\bibinfo {author} {\bibfnamefont {H.~M.}\ \bibnamefont
  {James}},\ }\href@noop {} {\bibfield  {journal} {\bibinfo  {journal} {Journal
  of Chemical Physics}\ }\textbf {\bibinfo {volume} {15}},\ \bibinfo {pages}
  {651} (\bibinfo {year} {1947})}\BibitemShut {NoStop}%
\bibitem [{\citenamefont {Edwards}(1969)}]{Edwards:1969dd}%
  \BibitemOpen
  \bibfield  {author} {\bibinfo {author} {\bibfnamefont {S.~F.}\ \bibnamefont
  {Edwards}},\ }\href@noop {} {\bibfield  {journal} {\bibinfo  {journal}
  {Journal of Physics Part C Solid State Physics}\ }\textbf {\bibinfo {volume}
  {2}},\ \bibinfo {pages} {1} (\bibinfo {year} {1969})}\BibitemShut {NoStop}%
\bibitem [{\citenamefont {Flory}(1985)}]{Flory:1985co}%
  \BibitemOpen
  \bibfield  {author} {\bibinfo {author} {\bibfnamefont {P.~J.}\ \bibnamefont
  {Flory}},\ }\href@noop {} {\bibfield  {journal} {\bibinfo  {journal} {Polymer
  Journal}\ }\textbf {\bibinfo {volume} {17}},\ \bibinfo {pages} {1} (\bibinfo
  {year} {1985})}\BibitemShut {NoStop}%
\bibitem [{\citenamefont {Holzl}, \citenamefont {Trautenberg},\ and\
  \citenamefont {Goritz}(1997)}]{Holzl:1997dd}%
  \BibitemOpen
  \bibfield  {author} {\bibinfo {author} {\bibfnamefont {T.}~\bibnamefont
  {Holzl}}, \bibinfo {author} {\bibfnamefont {H.~L.}\ \bibnamefont
  {Trautenberg}}, \ and\ \bibinfo {author} {\bibfnamefont {D.}~\bibnamefont
  {Goritz}},\ }\href@noop {} {\bibfield  {journal} {\bibinfo  {journal}
  {Physical Review Letters}\ }\textbf {\bibinfo {volume} {79}},\ \bibinfo
  {pages} {2293} (\bibinfo {year} {1997})}\BibitemShut {NoStop}%
\bibitem [{\citenamefont {de~Gennes}(1979)}]{DeGennes1979}%
  \BibitemOpen
  \bibfield  {author} {\bibinfo {author} {\bibfnamefont {P.-G.}\ \bibnamefont
  {de~Gennes}},\ }\href@noop {} {\emph {\bibinfo {title} {Scaling Concepts in
  Polymer Physics}}}\ (\bibinfo  {publisher} {Cornell University Press},\
  \bibinfo {year} {1979})\BibitemShut {NoStop}%
\bibitem [{\citenamefont {Rubinstein}\ and\ \citenamefont
  {Colby}(2003)}]{Rubinstein2003}%
  \BibitemOpen
  \bibfield  {author} {\bibinfo {author} {\bibfnamefont {M.}~\bibnamefont
  {Rubinstein}}\ and\ \bibinfo {author} {\bibfnamefont {R.~H.}\ \bibnamefont
  {Colby}},\ }\href@noop {} {\emph {\bibinfo {title} {Polymer Physics}}}\
  (\bibinfo  {publisher} {Oxford University Press},\ \bibinfo {year}
  {2003})\BibitemShut {NoStop}%
\bibitem [{\citenamefont {Stukowski}(2009)}]{Stukowski:2009ky}%
  \BibitemOpen
  \bibfield  {author} {\bibinfo {author} {\bibfnamefont {A.}~\bibnamefont
  {Stukowski}},\ }\href@noop {} {\bibfield  {journal} {\bibinfo  {journal}
  {Modelling and Simulation in Materials Science and Engineering}\ }\textbf
  {\bibinfo {volume} {18}},\ \bibinfo {pages} {015012} (\bibinfo {year}
  {2009})}\BibitemShut {NoStop}%
\bibitem [{\citenamefont {Anderson}, \citenamefont {Lorenz},\ and\
  \citenamefont {Travesset}(2008)}]{Anderson:2008bt}%
  \BibitemOpen
  \bibfield  {author} {\bibinfo {author} {\bibfnamefont {J.~A.}\ \bibnamefont
  {Anderson}}, \bibinfo {author} {\bibfnamefont {C.~D.}\ \bibnamefont
  {Lorenz}}, \ and\ \bibinfo {author} {\bibfnamefont {A.}~\bibnamefont
  {Travesset}},\ }\href@noop {} {\bibfield  {journal} {\bibinfo  {journal}
  {Journal of Computational Physics}\ }\textbf {\bibinfo {volume} {227}},\
  \bibinfo {pages} {5342} (\bibinfo {year} {2008})}\BibitemShut {NoStop}%
\bibitem [{\citenamefont {Glaser}\ \emph {et~al.}(2015)\citenamefont {Glaser},
  \citenamefont {Nguyen}, \citenamefont {Anderson}, \citenamefont {Lui},
  \citenamefont {Spiga}, \citenamefont {Millan}, \citenamefont {Morse},\ and\
  \citenamefont {Glotzer}}]{Glaser:2015cu}%
  \BibitemOpen
  \bibfield  {author} {\bibinfo {author} {\bibfnamefont {J.}~\bibnamefont
  {Glaser}}, \bibinfo {author} {\bibfnamefont {T.~D.}\ \bibnamefont {Nguyen}},
  \bibinfo {author} {\bibfnamefont {J.~A.}\ \bibnamefont {Anderson}}, \bibinfo
  {author} {\bibfnamefont {P.}~\bibnamefont {Lui}}, \bibinfo {author}
  {\bibfnamefont {F.}~\bibnamefont {Spiga}}, \bibinfo {author} {\bibfnamefont
  {J.~A.}\ \bibnamefont {Millan}}, \bibinfo {author} {\bibfnamefont {D.~C.}\
  \bibnamefont {Morse}}, \ and\ \bibinfo {author} {\bibfnamefont {S.~C.}\
  \bibnamefont {Glotzer}},\ }\href@noop {} {\bibfield  {journal} {\bibinfo
  {journal} {Computer Physics Communications}\ }\textbf {\bibinfo {volume}
  {192}},\ \bibinfo {pages} {97} (\bibinfo {year} {2015})}\BibitemShut
  {NoStop}%
\bibitem [{\citenamefont {Kindt}(2005)}]{Kindt:2005gb}%
  \BibitemOpen
  \bibfield  {author} {\bibinfo {author} {\bibfnamefont {J.~T.}\ \bibnamefont
  {Kindt}},\ }\href@noop {} {\bibfield  {journal} {\bibinfo  {journal} {Journal
  of Chemical Physics}\ }\textbf {\bibinfo {volume} {123}},\ \bibinfo {pages}
  {144901} (\bibinfo {year} {2005})}\BibitemShut {NoStop}%
\bibitem [{\citenamefont {Bell}(1978)}]{Bell:1978hj}%
  \BibitemOpen
  \bibfield  {author} {\bibinfo {author} {\bibfnamefont {G.}~\bibnamefont
  {Bell}},\ }\href@noop {} {\bibfield  {journal} {\bibinfo  {journal}
  {Science}\ }\textbf {\bibinfo {volume} {200}},\ \bibinfo {pages} {618}
  (\bibinfo {year} {1978})}\BibitemShut {NoStop}%
\bibitem [{\citenamefont {Evans}, \citenamefont {Berk},\ and\ \citenamefont
  {Leung}(1991)}]{Evans:1991hu}%
  \BibitemOpen
  \bibfield  {author} {\bibinfo {author} {\bibfnamefont {E.}~\bibnamefont
  {Evans}}, \bibinfo {author} {\bibfnamefont {D.}~\bibnamefont {Berk}}, \ and\
  \bibinfo {author} {\bibfnamefont {A.}~\bibnamefont {Leung}},\ }\href@noop {}
  {\bibfield  {journal} {\bibinfo  {journal} {Biophysical Journal}\ }\textbf
  {\bibinfo {volume} {59}},\ \bibinfo {pages} {838} (\bibinfo {year}
  {1991})}\BibitemShut {NoStop}%
\bibitem [{\citenamefont {Seifert}(2000)}]{Seifert:2000uk}%
  \BibitemOpen
  \bibfield  {author} {\bibinfo {author} {\bibfnamefont {U.}~\bibnamefont
  {Seifert}},\ }\href@noop {} {\bibfield  {journal} {\bibinfo  {journal}
  {Physical Review Letters}\ }\textbf {\bibinfo {volume} {84}},\ \bibinfo
  {pages} {2750} (\bibinfo {year} {2000})}\BibitemShut {NoStop}%
\bibitem [{\citenamefont {Tito}, \citenamefont {Storm},\ and\ \citenamefont
  {Ellenbroek}(2017)}]{Tito:2017db}%
  \BibitemOpen
  \bibfield  {author} {\bibinfo {author} {\bibfnamefont {N.~B.}\ \bibnamefont
  {Tito}}, \bibinfo {author} {\bibfnamefont {C.}~\bibnamefont {Storm}}, \ and\
  \bibinfo {author} {\bibfnamefont {W.~G.}\ \bibnamefont {Ellenbroek}},\
  }\href@noop {} {\bibfield  {journal} {\bibinfo  {journal} {Macromolecules}\
  }\textbf {\bibinfo {volume} {50}},\ \bibinfo {pages} {9788} (\bibinfo {year}
  {2017})}\BibitemShut {NoStop}%
\bibitem [{\citenamefont {Scheutjens}\ and\ \citenamefont
  {Fleer}(1979)}]{Scheutjens:1979ib}%
  \BibitemOpen
  \bibfield  {author} {\bibinfo {author} {\bibfnamefont {J.~M. H.~M.}\
  \bibnamefont {Scheutjens}}\ and\ \bibinfo {author} {\bibfnamefont {G.~J.}\
  \bibnamefont {Fleer}},\ }\href@noop {} {\bibfield  {journal} {\bibinfo
  {journal} {The Journal of Physical Chemistry}\ }\textbf {\bibinfo {volume}
  {83}},\ \bibinfo {pages} {1619} (\bibinfo {year} {1979})}\BibitemShut
  {NoStop}%
\bibitem [{\citenamefont {Vaca}\ \emph {et~al.}(2015)\citenamefont {Vaca},
  \citenamefont {Shlomovitz}, \citenamefont {Yang}, \citenamefont {Valentine},\
  and\ \citenamefont {Levine}}]{Vaca:2015cn}%
  \BibitemOpen
  \bibfield  {author} {\bibinfo {author} {\bibfnamefont {C.}~\bibnamefont
  {Vaca}}, \bibinfo {author} {\bibfnamefont {R.}~\bibnamefont {Shlomovitz}},
  \bibinfo {author} {\bibfnamefont {Y.}~\bibnamefont {Yang}}, \bibinfo {author}
  {\bibfnamefont {M.~T.}\ \bibnamefont {Valentine}}, \ and\ \bibinfo {author}
  {\bibfnamefont {A.~J.}\ \bibnamefont {Levine}},\ }\href@noop {} {\bibfield
  {journal} {\bibinfo  {journal} {Soft Matter}\ }\textbf {\bibinfo {volume}
  {11}},\ \bibinfo {pages} {4899} (\bibinfo {year} {2015})}\BibitemShut
  {NoStop}%
\bibitem [{\citenamefont {Mayumi}\ \emph {et~al.}(2013)\citenamefont {Mayumi},
  \citenamefont {Marcellan}, \citenamefont {Ducouret}, \citenamefont {Creton},\
  and\ \citenamefont {Narita}}]{Mayumi:2013gd}%
  \BibitemOpen
  \bibfield  {author} {\bibinfo {author} {\bibfnamefont {K.}~\bibnamefont
  {Mayumi}}, \bibinfo {author} {\bibfnamefont {A.}~\bibnamefont {Marcellan}},
  \bibinfo {author} {\bibfnamefont {G.}~\bibnamefont {Ducouret}}, \bibinfo
  {author} {\bibfnamefont {C.}~\bibnamefont {Creton}}, \ and\ \bibinfo {author}
  {\bibfnamefont {T.}~\bibnamefont {Narita}},\ }\href@noop {} {\bibfield
  {journal} {\bibinfo  {journal} {ACS Macro Letters}\ }\textbf {\bibinfo
  {volume} {2}},\ \bibinfo {pages} {1065} (\bibinfo {year} {2013})}\BibitemShut
  {NoStop}%
\bibitem [{\citenamefont {Yount}, \citenamefont {Loveless},\ and\ \citenamefont
  {Craig}(2005)}]{Yount:2005cv}%
  \BibitemOpen
  \bibfield  {author} {\bibinfo {author} {\bibfnamefont {W.~C.}\ \bibnamefont
  {Yount}}, \bibinfo {author} {\bibfnamefont {D.~M.}\ \bibnamefont {Loveless}},
  \ and\ \bibinfo {author} {\bibfnamefont {S.~L.}\ \bibnamefont {Craig}},\
  }\href@noop {} {\bibfield  {journal} {\bibinfo  {journal} {Angewandte Chemie
  International Edition}\ }\textbf {\bibinfo {volume} {44}},\ \bibinfo {pages}
  {2746} (\bibinfo {year} {2005})}\BibitemShut {NoStop}%
\end{thebibliography}%

\clearpage

\onecolumngrid

\section*{SUPPORTING INFORMATION}
Literature references below refer to the reference section of the main paper above.
\appendix

\begin{center}\rule{3in}{0.4pt}\end{center}

\section{Stretched Gaussian polymer binding}
\label{sec:StretchedGaussianPolymers}

In this section, we consider the binding distribution for adding one reversible crosslink between two Gaussian polymers having two of their endpoints fixed at the origin, while their other two endpoints are fixed at arbitrary coordinates.

Suppose that the first chain has its end fixed at $\mathbf{R}_1$, and the other at $\mathbf{R}_2$. The partition function for just one chain is
\begin{equation}
	Q^{\circ}(N_i, \mathbf{R}_i) = \exp{\left(-\frac{3 (x_i^2 + y_i^2 + z_i^2)}{2 N_i b^2}\right)}
\end{equation}
where $i$ is either $1$ or $2$. Equivalently, we can express this as the product of two connected subchains of length $n_i$ and $N_i - n_i$:
\begin{equation}
	Q^{\circ}(N_i, \mathbf{R}_i) = \frac{1}{b^3} \left(\frac{3N_i}{2 \pi n_i (N_i - n_i)} \right)^{3/2} \int_{\mathbf{R}'} w(\mathbf{R}') w'(\mathbf{R}_i | \mathbf{R}') \ d\mathbf{R}'.
\end{equation}
For both chains, the combined partition function is
\begin{align}
	Q_{free}(N_1, N_2, \mathbf{R}_1, \mathbf{R}_2) &= Q^{\circ}(N_1, \mathbf{R}_1) Q^{\circ}(N_2, \mathbf{R}_2) \\
	&= \exp{\left(-\frac{3 (x_1^2 + y_1^2 + z_1^2)}{2 N_1 b^2}\right)} \exp{\left(-\frac{3 (x_2^2 + y_2^2 + z_2^2)}{2 N_2 b^2}\right)} \\
	&= \exp{\left[-\frac{3}{2 b^2} \left(\frac{|\mathbf{R}_1|^2}{N_1} + \frac{|\mathbf{R}_2|^2}{N_2}\right)\right]}
\end{align}
The partition function for the two chains when we place a link at position $(n_1, n_2)$ is
\begin{align}
	Q_{bound}(n_1, n_2; N_1, N_2, \mathbf{R}_1, \mathbf{R}_2) = &\frac{1}{b^6} \left(\frac{3}{2 \pi}\right)^3 \left(\frac{N_1 N_2}{n_1 (N_1 - n_1) n_2 (N_2 - n_2)}\right)^{3/2} \nonumber \\
	& \times \int_{\mathbf{R}'} w_1(\mathbf{R}') w_1'(\mathbf{R}_1 | \mathbf{R}')  w_2(\mathbf{R}')  w_2'(\mathbf{R}_2 | \mathbf{R}') \ d\mathbf{R}'
\end{align}
Just focusing on $x$, we must evaluate
\begin{align}
	\int_{-\infty}^{\infty}{ \exp{\left\{-\frac{3}{2 b^2}  \left[\left(\frac{n_1 + n_2}{n_1 n_2}\right) x'^2 + \frac{(x_1 - x')^2}{N_1 - n_1} + \frac{(x_2 - x')^2}{N_2 - n_2}\right]\right\}} \ dx'}.
\end{align}
The integrand can be written as
\begin{align}
	&\exp{\left\{-\frac{3}{2 b^2}  \left[\left(\frac{n_1 + n_2}{n_1 n_2}\right) x'^2 + \frac{(x_1 - x')^2}{N_1 - n_1} + \frac{(x_2 - x')^2}{N_2 - n_2}\right]\right\}} \\
	&= \exp{\left\{-\frac{3}{2 b^2}  \left[\left(\frac{n_1 + n_2}{n_1 n_2}\right) x'^2 + \frac{x_1^2 - 2 x' x_1 + x'^2}{N_1 - n_1} + \frac{x_2^2 - 2 x' x_2 + x'^2}{N_2 - n_2}\right]\right\}}  \\
	&= \exp{\left[ -\frac{3}{2 b^2} \left( \frac{x_1^2}{N_1 - n_1} + \frac{x_2^2}{N_2 - n_2}\right)\right]} \nonumber \\
	& \times \exp{\left\{-\frac{3}{2 b^2}  \left[\left(\frac{n_1 + n_2}{n_1 n_2}\right) x'^2 + \frac{- 2 x' x_1 + x'^2}{N_1 - n_1} + \frac{- 2 x' x_2 + x'^2}{N_2 - n_2}\right]\right\}} 
\end{align}
The first factor depends only on $x_1$ and $x_2$, so it can be factored out of the integral over $x'$, leaving behind the seconed factor:
\begin{align}
	& \exp{\left\{-\frac{3}{2 b^2}  \left[\left(\frac{n_1 + n_2}{n_1 n_2}\right) x'^2 + \frac{- 2 x' x_1 + x'^2}{N_1 - n_1} + \frac{- 2 x' x_2 + x'^2}{N_2 - n_2}\right]\right\}} \\
	&= \exp{\left\{-\frac{3}{2 b^2} \left(\frac{n_1 + n_2}{n_1 n_2} + \frac{N_1 - n_1 + N_2 - n_2}{(N_1 - n_1)(N_2 - n_2)} \right) x'^2 + \frac{3}{b^2}  \left(\frac{x_1}{N_1 - n_1} + \frac{x_2}{N_2 - n_2} \right) x'\right\}} 
\end{align}
We must then integrate this over $x'$:
\begin{align}
	&\int_{-\infty}^{\infty} \exp{\left\{-\frac{3 \mathcal{A}}{2 b^2} x'^2 + \frac{3}{b^2}  \left(\frac{x_1}{N_1 - n_1} + \frac{x_2}{N_2 - n_2} \right) x'\right\}}  \ dx' \nonumber  \\
	&= \sqrt{\frac{\pi}{\mathcal{A}}} \exp{\left[ \frac{3}{2\mathcal{A} b^2} \left(\frac{x_1}{N_1 - n_1} + \frac{x_2}{N_2 - n_2} \right)^2 \right]} \nonumber \\
	&\mbox{ where } \mathcal{A} =  \left(\frac{1}{n_1} + \frac{1}{n_2} + \frac{1}{N_1 - n_1} + \frac{1}{N_2 - n_2} \right) 
\end{align}
Performing this integral over $y'$ and $z'$ leads to
\begin{align}
	& Q_{bound}(n_1, n_2; N_1, N_2, \mathbf{R}_1, \mathbf{R}_2) = \frac{1}{b^3} \left(\frac{3}{2 \pi}\right)^{3/2} \left(\frac{N_1 N_2}{\mathcal{A} n_1 (N_1 - n_1) n_2 (N_2 - n_2)}\right)^{3/2}  \nonumber \\
	& \times \exp{\left[ -\frac{3}{2 b^2} \left( \frac{| \mathbf{R}_1 |^2}{N_1 - n_1} + \frac{| \mathbf{R}_2 |^2}{N_2 - n_2}\right)\right]} \nonumber \\
	& \times \exp{\left\{ \frac{3}{2 \mathcal{A} b^2} \left[ \left(\frac{x_1}{N_1 - n_1} + \frac{x_2}{N_2 - n_2} \right)^2 + \left(\frac{y_1}{N_1 - n_1} + \frac{y_2}{N_2 - n_2} \right)^2 + \left(\frac{z_1}{N_1 - n_1} + \frac{z_2}{N_2 - n_2} \right)^2 \right] \right\}} \nonumber
\end{align}
To write the binding partition function in terms of $n = n_1 + n_2$, we must consider all permutations of binding positions $(n_1, n_2)$ that lead to $n$. For notational simplicity, we restrict to the case where $N_1 = N_2 = N$. Both $n_1$ and $n_2$ must be between $1$ and $N - 1$ (because segments $0$ and $N$ along the two polymers are already bound to permanent crosslinks). The total binding partition function for a linker bound to $n_1 + n_2 = n$ is the sum over all valid permutations of $(n_1, n_2)$:
\begin{equation}
	Q_{poly, b}(n; N, \mathbf{R}_1, \mathbf{R}_2) = \sum_{m = A}^{B}{Q_{bound}(m, n - m; N, N, \mathbf{R}_1, \mathbf{R}_2)}
\end{equation}
where
\begin{align}
	&A = \max{\left(n - N + 1, 1\right)} \\
	&B = \min{\left(n - 1, N - 1\right)}.
\end{align}
The binding free energy for attaching a reversible linker with $n_1 + n_2 = n$ is then computed as normal by $G_{poly,b}(n; N, \mathbf{R}_1, \mathbf{R}_2) = -kT \ln{Q_{poly, b}(n; N, \mathbf{R}_1, \mathbf{R}_2) }$.

\section{Molecular Dynamics Simulation Details}
\label{sec:MDDetails}

Our coarse-grained molecular dynamics simulation consists of two bead-spring polymers, each with $N = 100$ segments, bound together by a permanent crosslink bead at segment $50$. The position of the permanent crosslink is fixed to the origin of the simulation box, while the polymer chains are allowed to fluctuate. The box boundaries are periodic in all three dimensions, though the size of the box is set to be large enough that the polymer chains do not interact with their periodic images. Simulation parameters and quantities are all given in terms of fundamental model units of distance $\mathcal{D}$, energy $\mathcal{E}$ (taken to be the thermal unit $kT$), mass $\mathcal{M}$, and time $\tau=\sqrt{\mathcal{M}\mathcal{D}^2/\mathcal{E}}$. Calculations were carried out using the HOOMD-Blue molecular dynamics package (v2.1.1). \cite{Anderson:2008bt, Glaser:2015cu} The system contains no explicit or implicit solvent. All systems are integrated using a time step size of $dt = 0.001\tau$. 

The beads (segments) comprising the polymer chains are held together by strong harmonic bonds, with the bonding potential
\begin{equation}
	U_{bond}(r) = \frac{1}{2} k_{bond} (r - r_0)^2
\end{equation}
where $r$ is the separation distance between the two beads on a given timestep in the simulation, and $r_0$ is the bond rest length. We choose $k = 5000 \mathcal{E}/\mathcal{D}^2$, and $r_0 = 0.85 \mathcal{D}$. The bond between polymer segment $50$ on each chain, and the permanent crosslink bead, has $k = 1000 \mathcal{E}/\mathcal{D}^2$ with the same $r_0$. The ``helper linkers'' used in Figure \ref{fig:SimLocalStrength} (main text) are identical to the permanent crosslink bead, and are attached to their two neighbour polymers segments by the same type of harmonic bond. 

Each polymer segment (besides segment $50$) has a binding site attached to it via a harmonic bond with $k = 5000 \mathcal{E}/\mathcal{D}^2$, and $r_0 = 0.4 \mathcal{D}$ (henceforth called ``$r_b$''). The binding sites on the first polymer chain are distinguished as type ``A1'', while those on the second polymer chain are defined as ``A2''.

Reversible crosslink particles may also be added to the simulation box if desired. These particles are composed of a single bead with two antipodal binding sites, one of type ``B1'' and the other of type ``B2''. The binding sites are attached to their host bead by harmonic bonds with $k = 5000 \mathcal{E}/\mathcal{D}^2$, and $r_0 = 0.4 \mathcal{D}$. The two binding sites on each reversible linker are held at an angle of $\pi$ relative to each other by a strong angle potential of the form
\begin{equation}
	U_{angle}(\theta) = \frac{1}{2} k_{angle} (\theta - \pi)^2.
	\label{eqn:AnglePotential}
\end{equation}
Here $\theta$ is the angle between the two binding sites, $k_{angle} = 100 \mathcal{E}/\text{rad}^2$ is the strength of the angular potential, and $\pi$ is the rest angle.

\begin{table}
\centering
\begin{tabular}{ |p{3.0cm}|p{2.7cm}|p{3.0cm}|p{3.0cm}|p{2.7cm}|p{2.7cm}| } 
    \hline
    & Polymer Segment & Permanent Crosslink & Reversible Crosslink & Bind Sites A1/A2 & Bind Sites B1/B2 \\ \hline
    Polymer Segment & $\epsilon = 1, \sigma = 1$ & $\epsilon = 1, \sigma = 1$ & $\epsilon = 1, \sigma = 1$ & $\epsilon = 1, \sigma = R_{b} + 0.5$ & $\epsilon = 1, \sigma = R_{b} + 0.5$ \\ \hline
    Permanent Crosslink & & $\epsilon = 1, \sigma = 1$ & $\epsilon = 1, \sigma = 1$ & $\epsilon = 1, \sigma = R_{b} + 0.5$ & $\epsilon = 1, \sigma = R_{b} + 0.5$ \\ \hline
    Reversible Crosslink & & & $\epsilon = 1, \sigma = 1$ & $\epsilon = 1, \sigma = R_{b} + 0.5$ & $\epsilon = 1, \sigma = R_{b} + 0.5$ \\ \hline
    Bind Sites A1/A2 & & & & $\epsilon = 1, \sigma = 2 R_{b,rep}$ & (none) \\ \hline
    Bind Sites B1/B2 & & & & & $\epsilon = 1, \sigma = 2 R_{b,rep}$  \\
    \hline
\end{tabular}
\caption{Inverse power law potential parameters for the intermolecular interactions in the system. Units of measure are $\mathcal{E}$ for $\epsilon$, $\mathcal{D}$ for $\sigma$.}
\label{tbl:NickSuppLJParams}
\end{table}

All non-bonded polymer segments, binding sites, the permanent crosslink, and any reversible crosslinks interact with each other via a Lennard-Jones-like inverse power law potential,
\begin{align}
	U_{int}(r) &= 4 \epsilon \left(\frac{\sigma}{r}\right)^{12} \mbox{ for } r < r_{cut} \nonumber \\
	&= 0 \mbox{ otherwise}.
\end{align}
Here, $\epsilon$ is the strength of the potential, $\sigma$ is the width, and $r_{cut} = 3.0\mathcal{D}$ is the cut-off radius. The parameters of the potential for each pair of bead types in the system are given in Table \ref{tbl:NickSuppLJParams}. The effective radius of a binding sites is given by $R_b = 0.1 \mathcal{D}$.

Binding sites of type A repel each other with an effective radius of $R_{b,rep} = 0.45 \mathcal{D}$, and the same for B. However, a Gaussian binding attraction between binding sites of type A1 and B1, and A2 and B2, is defined by
\begin{align}
	U_{bind}(r) &= -\epsilon_{bind} \exp{\left[-\frac{1}{2} \left(\frac{r}{\sigma_{bind}}\right)^2\right]} \mbox{ for } r < r_{cut} \nonumber \\
	&= 0 \mbox{ otherwise}.
	\label{eqn:GaussianPot}
\end{align}
The binding strength is tuned by $\epsilon_{bind}$, and the binding range is set by the binding site radius, $\sigma_{bind} = 2 R_{b} \mathcal{D}$.

With these ingredients in place, reversible crosslinks can form one or two bonds with the polymer chains. The strength of binding is tuned by $\epsilon_{bind}$. The repulsion of binding sites of like type prevents, for example, two polymer segment binding sites from potentially attaching to a single reversible crosslink binding site. By virtue of separating the binding sites on the two polymers into two distinct categories, reversible crosslinks may also only form a link between a segment on polymer chain 1, and another segment on polymer chain 2.

For a given choice of $\epsilon_{bind}$, the \emph{effective} attraction strength between binding sites is actually weaker than $U_{bind}(r)$, because the binding sites are attached to their host beads. As noted in Table \ref{tbl:NickSuppLJParams}, the host beads interact with other host beads, and also non-bonded binding sites, via an inverse power law potential. Thus, the effective attraction strength between two binding sites is the sum of the Gaussian potential $U_{bind}(r)$, plus the sum of the repulsive contributions from the inverse power law potentials. The effective depth of the potential energy well for two binding sites can be calculated analytically by
\begin{equation}
	\epsilon_{\text{bind, eff}} = \min{\left[U_{bind}(r - 2 r_b) + U_{int,\text{host/host}}(r) + 2 U_{int,\text{host/binder}}(r - r_b)\right]}
\end{equation}
The three terms here correspond to, in order: the attractive Gaussian potential between the two binding sites; the inverse power law repulsion between the two host beads of the binding sites; and two factors of the inverse power law repulsion between a host and a binding site. The quantity $r_b$ is the length of the bond connecting a binding site to its host bead, noted above.

After initialisation, all systems are equilibrated in the NVT ensemble with Langevin dynamics for between $1 \times 10^7$ and $1 \times 10^8$ time steps. System statistics are then recorded within the same ensemble over $1 \times 10^8$ time steps (at intervals of $1 \times 10^4$ time steps).

\section{Lattice SCFT Network Details}
\label{sec:SCFTDetails}

Each bridge $i$ in the SCFT model is represented as a random walk with a fixed number of steps $N_i$ (which may be different for each bridge). The force along a bridge is approximated with the ideal chain model,
\begin{equation}
	F_i(L_i) =  \frac{2 \langle L_i \rangle k T}{N_i b^2},
	\label{eqn:BridgeForceSupp}
\end{equation}
where $b = 1$ is the width of a monomer (equal to the lattice unit size $1$ here), $\langle L_i \rangle$ is the average end-to-end distance of the bridge extracted directly from the SCFT calculation, and $kT$ is the energy unit (set to unity in the model). As $L_i$ of each bridge changes during each strain step, the force $F_i(L_i)$ of the bridge changes according to Eq. \ref{eqn:BridgeForceSupp}. The bridge is instantaneously and irreversibly cut when $F_i(L_i)$ exceeds $F^*$, a model parameter representing the threshold force for breaking the connection between a bridge and a node. The cut is performed randomly from either of the two nodes the bridge is attached to. The bridge then remains in the system, connected to the second node, for all subsequent strain steps.

The full protocol for carrying out a strain experiment in our coarse-grained SCFT model is as follows:
\begin{enumerate}
	\item{Define a polymer network. This consists of choosing: the number of bridges in the system, the number of segments in each bridge, how the bridge ends are connected together via nodes, and the initial width and height of the network.}
	\item{Use the lattice SCFT approach to approximate the equilibrium distribution of polymer conformations and node positions for the current system size.}
	\item{Compute the force $F_i$ along each bridge $i$ via Eq. \ref{eqn:BridgeForceSupp}.}
	\item{Cut each bridge that is bearing a force greater than $F^*$.}
	\item{Perform a strain step, in which the width of the network is increased by two lattice units ($2b$).}
	\item{Repeat Steps 2 - 5 until the desired final strain is reached.}
\end{enumerate}

A square network initially of $50 \times 50$ lattice units in size is employed in our calculations, with $283$ bridges and $190$ nodes. Of these nodes, $20$ are defined as left-hand boundary nodes, and another $20$ as right-hand boundary nodes; the remaining nodes are ``free''. The boundary nodes are spatially fixed, and initially placed at a width of $50$ lattice units apart from each other in the system. (Their positions in the vertical axis of the lattice are random.) They thus represent the scenario of the polymer network being suspended between two boundary plates. The total width of the system is varied by changing the spacing between the left-hand and right-hand boundary nodes. Note that the boundaries of the network are invisible to the polymer bridges and free nodes.

The polymer bridges are randomly connected to the boundary and free nodes in the system such that: each node has either two or three connections; and no two given nodes are connected by two bridges. In the system considered in the main text, only four out of the $190$ nodes have two connections, while the remaining nodes have three connections. The ``coordination number'' of three employed here is arbitrary, and can obviously be changed as desired.

The free nodes are assigned random initial positions within the $50 \times 50$ system, so that each bridge $i$ has an initial length $L_{i,\text{init}}$. The number of segments $N_i$ in each bridge is determined based on $L_{i,\text{init}}$, by enforcing that the initial tension in each bridge is $(2/3) kT / b$ using Eq. \ref{eqn:BridgeForce}. This initial tension can be changed or heterogenised, and represents a network pre-stress. For bridges where Eq. \ref{eqn:BridgeForce} yields an $N_i$ that is less than $30$ segments, the bridge length is set to $30$. As such, the network studied here has a heterogeneous pre-stress.

\end{document}